\documentclass[onecolumn,showpacs,aps,prd]{revtex4}

\usepackage{latexsym}
\usepackage{amsmath,amssymb}
\usepackage{tikz}

\newcommand{\lc}{\varepsilon}

\newcommand{\ds}{\displaystyle}

\newcommand{\celi}{\ensuremath{\mathbb{Z}}}
\newcommand{\realni}{\ensuremath{\mathbb{R}}}

\newcommand{\cA}{{\cal A}}
\newcommand{\cD}{{\cal D}}

\newcommand{\cM}{{\cal M}}

\newcommand{\cO}{{\cal O}}

\newcommand{\cT}{{\cal T}}

\begin{document}

\title{Causal Dynamical Triangulations in the Spincube Model of Quantum Gravity}

\author{Marko Vojinovi\'c}
 \email{vmarko@ipb.ac.rs}
\affiliation{Institute of Physics, University of Belgrade, Pregrevica 118, 11080 Belgrade, Serbia}


\begin{abstract}
We study the implications of the simplicity constraint in the spincube model of quantum gravity. By relating the edge lengths to the integer areas of triangles, the simplicity constraint imposes very strong restrictions between them, ultimately leading to a requirement that all 4-simplices in the triangulation must be almost mutually identical. As a surprising and unexpected consequence of this property, one can obtain the CDT state sum as a special case of the spincube state sum. This relationship brings new insight into the long-standing problem of the relationship between the spinfoam approach and the CDT approach to quantum gravity. In particular, it turns out that the spincube model contains properties of both approaches, providing a single unifying framework for their analysis and comparison. In addition, the spincube state sum also contains some other special cases, very similar but not equivalent to the CDT state sum.
\end{abstract}

\maketitle

\section{\label{SecIntroduction}Introduction}

Loop Quantum Gravity (LQG) \cite{AstekarPrviRad,RovelliBook} is one of the mainstream approaches to the fundamental problem of the gravitational field quantization. Various formulations of concrete models of LQG can be roughly divided into canonical and covariant. The canonical formulation focuses on the Hamiltonian formalism of gravity and its canonical quantization. The covariant formulation focuses on providing a rigorous definition to the gravitational path integral
$$
Z = \int \cD g_{\mu\nu}\, e^{iS[g]}\, .
$$
This is mostly being done by discretizing the integral, like in ordinary quantum mechanics. The models thus obtained are generically called state sum models (see Appendix \ref{ApendiksA}), and they describe gravitational and matter fields by suitably labeling a piecewise-linear $4$-manifold. In other words, the path integral is defined as a sum over certain labels on a $4$-dimensional lattice, most commonly an irregular triangulation. The most developed models of this type are called spinfoam models, see for example \cite{EPRL,FK,RovelliVidottoBook}. An alternative definition of the path integral is to fix the labels to have constant values, and perform the sum over some class of manifold triangulations instead. The most developed model of this type is called Causal Dynamical Triangulations approach (CDT), see \cite{CDTreview} for a review. As it turns out, the two choices of what to sum over in the path integral lead to related but very different models, whose properties are not easy to compare. This represents a long-standing open problem of the relationship between the spinfoam approaches and the CDT approach.

Recently, a novel generalization of spinfoam models has appeared, called the {\it spincube} model \cite{MikovicVojinovic2012}. It represents a generalization of spinfoams through higher category theory \cite{BaezHuerta2011}. In particular, whereas spinfoam models are based on the structure of a Lie group, the spincube model is based on the structure of a Lie $2$-group, a $2$-categorical generalization of the notion of a group (see \cite{BaezHuerta2011} for a comprehensive list of references on $2$-groups). As a consequence, while in spinfoam models one uses the structure of a labeled simplicial $2$-complex, in the spincube model one uses the structure of a labeled simplicial $3$-complex (hence the names ``spinfoam'' and ``spincube''). The spincube model was introduced in order to improve certain properties of spinfoam models, like the presence of the tetrad fields, coupling of matter fields to gravity and the semiclassical limit, which are issues hard to deal with in spinfoam models.

It thus comes as a fresh surprise that the spincube model naturally incorporates both the CDT and spinfoam approaches within a single model. In this paper we will demonstrate explicitly how and why this happens, thus giving some new insight to the relationship between spinfoams and CDT discussed above. This is an unexpected result, which came about accidentally, during the study of certain properties of the spincube model.

In short, what happens is the following. The spincube model consists of the state sum quantization of a topological $BFCG$ theory \cite{GirelliPfeifferPopescu2008,FariaMartinsMikovic2011}, to which one then applies a simplicity constraint. This is in complete analogy with spinfoam models, which are constructed by starting from a state sum of the topological $BF$ theory and then applying the simplicity constraint. As it turns out, in the spincube model the simplicity constraint has a simple geometric interpretation --- the areas of all triangles (as calculated using the edge lengths for each given triangle) must have integer values, in certain units. But since there are more triangles than edges in any given triangulation, the simplicity constraint is represented by a system of equations which is overcomplete in edge lengths, while taking into account integer areas as variables leads to a complicated system of Diophantine equations, which may happen to have no solutions at all (this was first noticed in \cite{MikovicStrongWeak}). A more detailed analysis of the simplicity constraint equations was thus necessary, to study whether they can be imposed at all or not. That analysis is presented in this paper, with a result that is doubly surprising. First, despite the complicated Diophantine equations, the set of solutions for the simplicity constraint is indeed nonempty, and several classes of solutions will be explicitly constructed. And second, all these solutions have a nontrivial resemblance to the CDT model, while one class turns out to transform the spincube state sum model precisely into the CDT state sum model.

One can thus say that the spincube model represents a generalization of both spinfoam and CDT approaches, and encapsulates the properties of both into a single model. This is very interesting, as it opens novel avenues for the study of both approaches, and especially their mutual comparison.

The layout of the paper is as follows. In section \ref{SecSpincubeModel} we will give a short overview of the spincube model of quantum gravity, introduce notation and conventions. An important step in the construction of the model is the way one imposes the simplicity constraint. Section \ref{SecImposingTheSimplicityConstraint} is devoted to a detailed discussion of the difference between imposing the simplicity constraint weakly and strongly. It will turn out that the strong simplicity constraint is nontrivial to impose, and might not have any solutions. In section \ref{SecSolutionToTheSimplicityConstraint} we give a proof that the strong simplicity constraint actually does have solutions, and we construct one class. Another class of solutions is then discussed in section \ref{SecRelationToCDT}, and we show that for these solutions the spincube model reduces precisely to the CDT model of quantum gravity. Section \ref{SecTopology} deals with the implications of possible nontrivial spacetime topologies to the spincube model. Our conclusions and discussion of the results are given in section \ref{SecConclusions}, while the Appendix contains some further technical information.

The notation and conventions are as follows. The local Lorentz indices are denoted by the Latin letters $a,b,c,\dots$, take values $0,1,2,3$, and are raised and lowered using the Minkowski metric $\eta_{ab}$ with signature $(-,+,+,+)$. Spacetime indices are denoted by the Greek letters $\mu,\nu,\dots$, and are raised and lowered by the spacetime metric $g_{\mu\nu} = \eta_{ab} e^a{}_{\mu} e^b{}_{\nu}$, where $e^a{}_{\mu}$ are the tetrad fields.

\section{\label{SecSpincubeModel}Spincube Model}

We begin by giving a review of the \textit{spincube model}, which was introduced in detail in \cite{MikovicVojinovic2012}. In this paper we do not aim to provide a detailed  motivation for the introduction of the spincube model, nor a detailed analysis. Rather, we merely repeat the minimum self-contained amount of material given in \cite{MikovicVojinovic2012} needed to fix the notation and to spell out the main properties of the model. A reader interested in more details about the model itself, the Poincare $2$-group and representation theory of $2$-categories can look up the literature \cite{MikovicVojinovic2012,BaezHuerta2011,GirelliPfeifferPopescu2008,FariaMartinsMikovic2011,BaezBaratinFreidelWise2012,BaratinFreidel2015}.

Let $\cM_4$ be a four-dimensional spacetime manifold. The spincube model of quantum gravity is based on the classical action
\begin{equation} \label{SpincubeDejstvo}
S = \int_{\cM_4} B^{ab}\wedge R_{ab} + e^a\wedge G_a - \phi^{ab} \wedge \left( B_{ab} - \lc_{abcd}\, e^c\wedge e^d \right),
\end{equation}
which is called the constrained $BFCG$ action for the Poincar\'e $2$-group. The first two terms represent the topological $BFCG$ action \cite{GirelliPfeifferPopescu2008,FariaMartinsMikovic2011}, which is a $2$-categorical generalization of the standard $BF$ action. The last term is called the simplicity constraint. The fundamental variables in this action are two-forms $B^{ab}$, $\phi^{ab}$, $\beta^a$ and one-forms $e^a$, $\omega^{ab}$. The $B^{ab}$, $\phi^{ab}$ and $e^a$ are Lagrange multipliers, while $\omega^{ab}$ and $\beta^a$ together constitute the $2$-connection for the Poincar\'e $2$-group. They give rise to corresponding two-form and three-form field strengths
$$
R^{ab} \equiv d\omega^{ab} + \omega^a{}_c \wedge \omega^{cb}, \qquad G^a \equiv d\beta^a + \omega^a{}_c\wedge \beta^c.
$$
Note that the quantities traditionally denoted $F_{ab}$ and $C^a$ in the $BFCG$ action have been conveniently relabeled here as $R_{ab}$ and $e^a$, since they have the interpretation of spacetime curvature two-form and the tetrad one-form, respectively.

The variation of the action (\ref{SpincubeDejstvo}) with respect to all of its variables yields the following set of equations of motion:
$$
\begin{array}{lccl}
\delta B^{ab} &: && R_{ab}-\phi_{ab} = 0 , \\
\delta \phi^{ab} &: && B_{ab} - \lc_{abcd}\, e^c\wedge e^d = 0 , \\
\delta \omega^{ab} &: &&  \nabla B_{ab} - e_{[a} \wedge \beta_{b]} = 0, \\
\delta \beta^a &: && \nabla e_a = 0 , \\
\delta e^a &: && G_a +2\lc_{abcd}\, \phi^{bc}\wedge e^d = 0 . \\
\end{array}
$$
If the tetrad field is assumed to be nondegenerate, $\det ( e^a{}_{\mu}) \neq 0$, this system of equations can be rewritten into an equivalent system, consisting of the equations that solve for the Lagrange multipliers,
$$
\phi_{ab} = R_{ab}, \qquad B_{ab} = \lc_{abcd}\, e^c\wedge e^d,
$$
the equations for $\beta^a$ and torsion two-form $T^a \equiv \nabla e^a = de^a + \omega^a{}_c \wedge e^c$,
$$
\beta^a = 0, \qquad T^a = 0,
$$
and the equation for the tetrad field $e^a$,
$$
\lc_{abcd} \, R^{bc}\wedge e^d = 0,
$$
which is the differential-form notation for the Einstein equations of general relativity. Thus we see that the action (\ref{SpincubeDejstvo}) is a reformulation of general relativity (or more precisely of the Einstein-Cartan theory, see \cite{MikovicVojinovic2012}).

This reformulation is convenient because it has two key properties. First, the action is split into the topological $BFCG$ part and the simplicity constraint. This is in complete analogy with the Plebanski action, which is split into the topological $BF$ part and the simplicity constraint. The usefulness of this split lies in the fact that one can perform the quantization of the topological sector, and then impose the simplicity constraint, thus defining a theory of quantum gravity. In the case of the Plebanski action, this program leads to spinfoam models of quantum gravity (see \cite{EPRL,FK,RovelliVidottoBook}). In the case of the action (\ref{SpincubeDejstvo}), the same program leads to a $2$-categorical generalization of spinfoam models, dubbed {\it spincube} model. The difference lies only in the fact that the quantization of $BF$ theory is based on the underlying structure of a group (giving rise to a colored $2$-complex, hence ``spinfoam''), while the quantization of the $BFCG$ theory is based on the underlying structure of a $2$-group (giving rise to the colored $3$-complex, hence ``spincube'').

The second key property is that, in contrast to the Plebanski action, the action (\ref{SpincubeDejstvo}) features tetrad fields as explicit variables in the topological sector. After the quantization of the topological sector, the presence of tetrads manifests itself as the presence of a new set of colors on the $3$-complex, which are interpreted as edge lengths. The spinfoam models do not feature these labels, and consequently run into trouble when one tries to couple matter fields, especially massive fermions (see Appendix in \cite{BHMPRW2013} for a short overview). In contrast, the coupling of any type of matter fields to the spincube model is completely straightforward \cite{MikovicVojinovic2012}, and represents a clear improvement over the spinfoam models.

Specifically, the mass terms of matter fields in the discretized action (and also the cosmological constant term) depend on the $4$-volume of each $4$-simplex in the triangulation. This $4$-volume term may be hard or even impossible to express in terms of the fundamental variables of a given spinfoam model. For example, in area-Regge variables the $4$-simplex is assigned $10$ areas for its triangles. These variables do not fix a unique value of the $4$-volume (see \cite{BarrettRocekWilliams1999} and also Appendix \ref{ApendiksD}), so mass terms in the discretized action are not well-defined. Similarly, in the Engle-Pereira-Rovelli-Livine/Freidel-Krasnov spinfoam model one uses triangle areas and tetrahedra $3$-volumes as variables, which also do not fix a unique choice of the $4$-volume without the knowledge of at least one edge length \cite{BHMPRW2013}, which is not a variable of the theory. In the spincube model, this issue is resolved in arguably the simplest possible and geometrically most natural way, due to the explicit presence of tetrad fields in the topological part of the action (\ref{SpincubeDejstvo}). On a triangulation, these variables become the edge lengths, so given a set of $10$ edge lengths in a $4$-simplex, its $4$-volume is uniquely determined and easily expressed in terms of the Cayley-Menger determinant formula (see (\ref{CayleyDet})). Therefore, all mass terms in the discretized action are automatically well-defined.

Let us now summarize the details of the quantum theory corresponding to (\ref{SpincubeDejstvo}). The spincube model is one particular example of a state sum model (see Appendix \ref{ApendiksA} for a short introduction to state sum models), and is defined as follows. Let $T(\cM_4)$ be a triangulation of the spacetime manifold $\cM_4$. Triangulation $T$ contains vertices $v$, edges $\epsilon$, triangles $\Delta$, tetrahedra $\tau$ and $4$-simplices $\sigma$. The total numbers of edges and triangles will be denoted as $E$ and $F$, respectively. Each edge $\epsilon$ is labeled by a non-negative parameter $l_{\epsilon} \in \realni_0^+$, each triangle $\Delta$ is labeled by an integer $m_{\Delta}\in\celi$, while vertices, tetrahedra and $4$-simplices do not carry any additional labels. Thus the set of all $l_{\epsilon}$, $m_{\Delta}$ represents the fundamental variables which describe the gravitational field. Further, the vertex amplitude $\cA_v$ and the tetrahedron amplitude $\cA_{\tau}$ are chosen to be trivial,
$$
\cA_v (l,m) = 1\,, \qquad \cA_{\tau}(l,m) =1\,,
$$
 while the $4$-simplex amplitude is chosen as
$$
\cA_{\sigma} = e^{ iS_{\sigma}(l,m) }\,,
$$
where $S_{\sigma}$ is the Regge-like action for a given $4$-simplex $\sigma$,
$$
S_{\sigma} (l,m) = \sum_{\Delta\in\sigma} |m_{\Delta}| \Theta_{\Delta,\sigma}(l)\,.
$$
Here $\Theta_{\Delta,\sigma}(l)$ is the dihedral angle corresponding to the triangle $\Delta$ of the $4$-simplex $\sigma$, and is a function of the edge lengths $l_{\epsilon}$ of $\sigma$. As we shall discuss in the next section, the integer $m_{\Delta}$ is related to the area of the triangle $\Delta$.

The state sum of the spincube model is then written as
\begin{equation}
Z_{T} = \int_{\realni_0^+} dl_1 \dots \int_{\realni_0^+} dl_E \sum_{m_1\in \celi} \dots \sum_{m_F\in\celi} \prod_{\epsilon\in T} \cA_{\epsilon}(l,m) \prod_{\Delta\in T} \cA_{\Delta}(l,m) \prod_{\sigma\in T} e^{iS_{\sigma}(l,m)} \,.
\end{equation}
The product over all $4$-simplices can be rewritten in the form
\begin{equation} \label{ReggeActionRacun}
\begin{array}{ccl}
\ds \prod_{\sigma\in T} e^{iS_{\sigma}(l,m)} & = & \ds \exp \left[i \sum_{\sigma\in T} S_{\sigma}(l,m)\right] \\
 & = & \ds \exp \left[ i \sum_{\sigma\in T} \sum_{\Delta\in\sigma} |m_{\Delta}| \Theta_{\Delta,\sigma}(l) \right] \\
 & = & \ds \exp \left[ i \sum_{\Delta\in T} |m_{\Delta}| \left(\sum_{\sigma\ni \Delta} \Theta_{\Delta,\sigma}(l) \right) \right] \\
 & = & \ds \exp \left[ i \sum_{\Delta\in T} |m_{\Delta}| \delta_{\Delta}(l) \right] \\
 & = & \ds  e^{ i S_R(l,m) }\, , \vphantom{\left[ \sum_{\Delta} \right]} \\
\end{array}
\end{equation}
where $\delta_{\Delta} = \sum_{\sigma\ni\Delta} \Theta_{\Delta,\sigma}$ is the deficit angle for the triangle $\Delta$ in Minkowski space, and $S_R$ is the total Regge-like action over the triangulation $T(\cM_4)$. It differs from the usual Regge action by the presence of the integer $m_{\Delta}$ in place of the area of the triangle, $A_{\Delta}(l)$ --- the former is an integer independent of edge lengths, while the latter is a function of edge lengths given by the Heron formula. This difference will be discussed in detail in the next section.

The spincube state sum can thus be written as
\begin{equation} \label{SpincubeStateSuma}
Z_{T} = \int_{\realni_0^+} dl_1 \dots \int_{\realni_0^+} dl_E \sum_{m_1\in \celi} \dots \sum_{m_F\in\celi} \prod_{\epsilon\in T} \cA_{\epsilon}(l,m) \prod_{\Delta\in T} \cA_{\Delta}(l,m) \; e^{iS_R(l,m)} \,,
\end{equation}
where the remaining edge and triangle amplitudes $\cA_{\epsilon}$ and $\cA_{\Delta}$ are yet to be specified. Their forms are based on the way we choose to impose the simplicity constraint, which is the main topic of the next section.

One should note that the model can be easily extended to include matter fields, cosmological constant, etc. However, in this paper we will omit such extensions in order to make the exposition more transparent.

\section{\label{SecImposingTheSimplicityConstraint}Imposing the Simplicity Constraint}

Traditionally, given a classical action, one constructs the state sum model in the following way. First, one rewrites the classical action into a topological sector and the simplicity constraint. Second, one constructs the state sum model for the topological sector of the theory, so that the resulting state sum is a triangulation independent topological invariant. Third, one modifies the amplitudes and imposes the simplicity constraint in a certain way, in order to pass from the topological state sum model to the non-topological state sum model which represents the quantization of the full classical action.

The crucial difference between the topological and non-topological state sum models lies in the implementation of the simplicity constraint, which represents the difference between the full theory and its topological sector. The simplicity constraint is implemented as a suitable relation connecting the independent variables of the topological state sum model. In the case of the spincube model, those variables are positive real numbers $l_{\epsilon}$ and integers $m_{\Delta}$, associated to edges and triangles in the triangulation, respectively. In the variables of the classical action (\ref{SpincubeDejstvo}) the simplicity constraint reads
$$
B^{ab} = \lc^{abcd}\, e_c \wedge e_d \,,
$$
and is a classical equation of motion obtained by varying the action (\ref{SpincubeDejstvo}) with respect to $\phi_{ab}$. Passing to the variables $l_{\epsilon},m_{\Delta}$ defined on the triangulation, the simplicity constraint takes the following form:
\begin{equation} \label{SimplicitySistemJna}
 \gamma l_p^2 \, | m_{\Delta} | = A_H(l_{\epsilon_1},l_{\epsilon_2},l_{\epsilon_3})\,, \qquad \forall \Delta\in T(\cM_4) \,,
\end{equation}
where $\epsilon_1,\epsilon_2,\epsilon_3\in\Delta$. Here $A_H(l)$ is the Heron formula for the area of the triangle $\Delta$ with edge lengths $l$,
\begin{equation} \label{HeronovaFormula}
A_H(l_1,l_2,l_3) = \sqrt{s(s-l_1)(s-l_2)(s-l_3)}\,,
\end{equation}
where $s \equiv (l_1+l_2+l_3)/2$, while $\gamma$ is a dimensionless parameter determining the overall scale, and the unit of length is provided by the Planck length $l_p$. The simplicity constraint has a natural and obvious geometric interpretation --- for every triangle $\Delta$ in $T$, the integer $m_{\Delta}$ determines the area of the triangle, with a universal proportionality constant $\gamma l_p^2$.

We should note that the factor $\gamma$ is a free parameter of the theory. It should not be confused with the Barbero-Immirzi parameter which is also commonly denoted by the same letter $\gamma$. The Barbero-Immirzi parameter can be introduced as a coupling constant if one adds the so-called Holst term \cite{Holst1996} to the action (\ref{SpincubeDejstvo}). In this paper we do not introduce it.

One can immediately see two basic properties of the simplicity constraint. First, it imposes triangle inequalities on all edge lengths $l_{\epsilon}$ in the triangulation. Namely, in triangulations with Lorentzian signature it is perfectly possible for the three edge lengths to violate triangle inequalities. This would give rise to an imaginary value of area (\ref{HeronovaFormula}), which corresponds to a triangle coplanar with some time-axis in Minkowski spacetime. However, simplicity constraint (\ref{SimplicitySistemJna}) requires all triangles to be ``spacelike'', i.e. to have real-valued (and positive) areas, which in turn imposes the triangle inequalities through (\ref{HeronovaFormula}).

Second, the simplicity constraint transforms the Regge-like action from (\ref{ReggeActionRacun}),
$$
S_R(l,m) = \sum_{\Delta\in T} |m_{\Delta}| \delta_{\Delta}(l)
$$
into the proper Regge action
\begin{equation} \label{ProperReggeAction}
S_{\rm Regge}(l) = \frac{1}{\gamma l_p^2} \sum_{\Delta\in T} A_{\Delta}(l)\delta_{\Delta}(l) \,.
\end{equation}

However, there is one big issue with the simplicity constraint (\ref{SimplicitySistemJna}). Namely, as it was first noted in \cite{MikovicStrongWeak},  the system of equations (\ref{SimplicitySistemJna}) is not guaranteed to have any solutions. This can be seen as follows. If the total number of edges in $T$ is $E$, and the total number of triangles is $F$, then for every triangulation we have $F\geq E$, while the equality holds only for a single $4$-simplex. This means that we have in total $F$ equations for $F$ integer variables and $E$ real variables. If we write the system generically as
$$
\begin{array}{lcl}
|m_1| & = & A_1(l_1,\dots,l_E)\, , \\
 & \vdots & \\
|m_E| & = & A_E(l_1,\dots,l_E)\, , \\
|m_{E+1}| & = & A_{E+1}(l_1,\dots,l_E)\,, \\
 & \vdots & \\
|m_F| & = & A_F(l_1,\dots,l_E)\,, \\
\end{array}
$$
where $A_1,\dots,A_F$ are suitable functions, we can in principle solve the first $E$ equations for $l_{\epsilon}$ as functions of $m_1,\dots,m_E$. Substituting those expressions into the remaining $F-E$ equations, the latter can be written in the form
\begin{equation} \label{OpstiSistemDiofantovihJna}
\begin{array}{lcl}
|m_{E+1}| & = & f_1(m_1,\dots,m_E)\,, \\
 & \vdots & \\
|m_{F}| & = & f_{F-E}(m_1,\dots,m_E)\,, \\
\end{array}
\end{equation}
where $f_1,\dots,f_{F-E}$ are implicitly defined functions, too complicated to be expressible in terms of elementary functions. Given that all $m_{\Delta}$ are integers, equations (\ref{OpstiSistemDiofantovihJna}) are thus very complicated Diophantine equations, and they are not guaranteed to have any solutions at all. Therefore, short of providing the proof that the set of solutions for (\ref{OpstiSistemDiofantovihJna}) is never empty, the simplicity constraint system (\ref{SimplicitySistemJna}) may be impossible to enforce as it stands.

In \cite{MikovicStrongWeak} an alternative strategy was proposed --- to implement the constraint ``weakly'', such that it holds only in the classical limit of the theory. In particular, it should be written as
\begin{equation} \label{SimplicitySistemJnaMikovic}
|m_{\Delta}| = \left\lfloor \frac{1}{\gamma l_p^2} A_H(l_{\epsilon_1},l_{\epsilon_2},l_{\epsilon_3}) \right\rfloor\,, \qquad \forall \Delta\in T(\cM_4) \,,
\end{equation}
where again $\epsilon_1,\epsilon_2,\epsilon_3\in\Delta$. Here $\lfloor \dots \rfloor$ represents the ``floor'' function, which returns the integer part of its argument. This definition of the simplicity constraint completely circumvents the issue of Diophantine equations. Moreover, as argued in \cite{MikovicStrongWeak}, in the classical limit (defined as an asymptotic expansion when $m_{\Delta},l_{\epsilon}\to\infty$) the difference between (\ref{SimplicitySistemJna}) and (\ref{SimplicitySistemJnaMikovic}) is always a lower order correction, making the two systems of equations asymptotically equivalent.

In the formalism of the state sum model, the implementation of the weak simplicity constraint (\ref{SimplicitySistemJnaMikovic}) is given by the following choice of the edge and triangle amplitudes,
$$
\cA_{\epsilon}(l,m) =  1\,, \qquad
\cA_{\Delta}(l,m) =  \chi \left( |m_{\Delta}| - \left\lfloor \frac{1}{\gamma l_p^2} A_H(l_{\epsilon\in\Delta}) \right\rfloor \right) \,, 
$$
where $\chi(x)$ is an indicator function, having value $1$ if $x=0$ and is zero otherwise. Substituting this into the state sum (\ref{SpincubeStateSuma}), one can immediately perform the summations over all $m_{\Delta}$, and the state sum reduces to the one defining the Regge quantum gravity model:
\begin{equation} \label{ReggeQGstateSuma}
Z_{T}^{\rm weak} = \int dl_1 \dots dl_E \; e^{iS_R(l)} \,,
\end{equation}
where the domain of integration over edge lengths is a complicated subset of $(\realni_0^+)^E$ due to the triangle inequalities imposed by the simplicity constraint. The action in the exponent is
\begin{equation} \label{ReggeDejstvoSaWeakConstraints}
S_R(l) \equiv \sum_{\Delta\in T} \left\lfloor \frac{1}{\gamma l_p^2} A_H(l_{\epsilon\in\Delta}) \right\rfloor \delta_{\Delta}(l)\,,
\end{equation}
which is asymptotically equal to the proper Regge action (\ref{ProperReggeAction}) in the classical limit $l_{\epsilon}\to k l_{\epsilon}$, $k\to\infty$ (see Appendix \ref{ApendiksC} for proof). Note that the choice of the edge and triangle amplitudes defines the measure of the discretized path integral, and the product of these amplitudes over all edges and triangles can be identified with an appropriate Jacobian determinant.

Nevertheless, one can argue that imposing the simplicity constraint weakly might be unsatisfactory on various grounds, and it is legitimate to ask if the constraint can be imposed strongly. Under the assumption that the Diophantine system discussed above has at least one solution for $l_{\epsilon}$, denoted $l_{\epsilon}=L_{\epsilon}(\bar{m})$, where the $\bar{m}$ denote the set of variables unconstrained by the equations, one can implement the strong simplicity constraint (\ref{SimplicitySistemJna}) with the following choice of edge and triangle amplitudes:
$$
\cA_{\epsilon}(l,m) = \delta\left(l_{\epsilon}- L_{\epsilon}(\bar{m})\right)\,, \qquad
\cA_{\Delta}(l,m) = \chi \left( |m_{\Delta}| - \frac{1}{\gamma l_p^2} A_H\left(L_{\epsilon\in\Delta}(\bar{m})\right) \right) \,.
$$
Substituting into (\ref{SpincubeStateSuma}), we can now integrate over all edge lengths, and evaluate all sums over $m$ except those over $\bar{m}$. The resulting state sum is
\begin{equation} \label{SpincubeStateSumaStrong}
Z_{T}^{\rm strong} = \sum_{ \{\bar{m}\} } \; e^{iS_R(L(\bar{m}))} \,,
\end{equation}
where the sum is taken over all independent sets of integers $\bar{m}$. The action in the exponent is
$$
S_R(L(\bar{m})) \equiv \frac{1}{\gamma l_p^2} \sum_{\Delta\in T} A_H(L_{\epsilon\in\Delta}(\bar{m})) \delta_{\Delta}(L(\bar{m}))\,,
$$
and it is equal to the proper Regge action (\ref{ProperReggeAction}) evaluated on the simplicity constraint solution $l_{\epsilon}=L_{\epsilon}(\bar{m})$.

The implementation of the strong simplicity constraint rests upon the assumption that the complicated Diophantine equations (\ref{OpstiSistemDiofantovihJna}) have a nonempty set of solutions, which is not obvious. It may thus come as a fresh surprise that a few classes of solutions can indeed be found. Moreover, the resulting spincube model can then be naturally related to a completely independent approach to quantum gravity, namely the Causal Dynamical Triangulations approach \cite{CDTreview}. This is both completely unexpected and a very interesting result.

In the next section, we will perform an explicit construction of one class of exact solutions of the strongly imposed simplicity constraint, and after that we will discuss the relation between the spincube model and the CDT approach to quantum gravity.

\section{\label{SecSolutionToTheSimplicityConstraint}Solution of the Simplicity Constraint}

Let us turn to a constructive proof that the simplicity constraint (\ref{SimplicitySistemJna}) always has at least one solution. The proof will be done in three steps. First we will discuss the case of a single $4$-simplex, then the case of two $4$-simplices sharing a common tetrahedron, and finally the general case of an arbitrary triangulation.

A single $4$-simplex has $10$ edges and $10$ triangles. In that case, the simplicity constraint (\ref{SimplicitySistemJna}) has the general form:
\begin{equation} \label{SistemJnaZaJedanSimplex}
\begin{array}{lcl}
 | m_1 | & = & \ds \frac{1}{ \gamma l_p^2} A_1(l_1,\dots,l_{10})\, , \\
 & \vdots & \\
 | m_{10} | & = & \ds \frac{1}{ \gamma l_p^2} A_{10}(l_1,\dots,l_{10})\, . \\
\end{array}
\end{equation}
This system of equations always has solutions. A simplest example is an equilateral $4$-simplex, with areas and edges given as
$$
m_1 = \dots = m_{10} = k\,, \quad l_1 = \dots = l_{10} = 2 l_p \sqrt{\frac{\gamma |k|}{\sqrt{3}}}\,,
$$
for any $k\in\celi$. For more complicated choices of $m$'s, numerical analysis suggests that several solutions for $l$'s may exist, since there are ten polynomial equations to be solved. However, note that we require the edge lengths to be real-valued and positive, and moreover they must satisfy triangle inequalities. The choice of $m$'s might be such that this is impossible to satisfy, in which case the system does not have any solutions. Therefore, it is important to stress that we are not attempting to solve the system for ten $l$'s given any arbitrary $m$'s, but rather to solve the system for $l$'s and for $m$'s simultaneously, within their respective domains. The equilateral example proves that the set of solutions is nonempty, and numerical analysis (of Monte-Carlo type) shows that the set of solutions is actually quite rich (see Appendix \ref{ApendiksD} for details).

Next we move to a less trivial case of two $4$-simplices sharing a common tetrahedron. The $4$-dimensional figure is depicted on the following diagram:
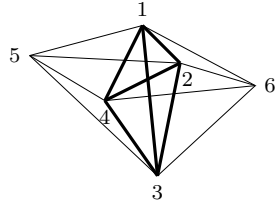
\begin{center}
\begin{tikzpicture}
\draw[very thick] (1,2) -- (1.5,1.5) ;
\draw[very thick] (1.5,1.5) -- (1.2,0) ;
\draw[very thick] (1.2,0) -- (0.5,1) ;
\draw[very thick] (0.5,1) -- (1,2) ;
\draw[very thick] (1,2) -- (1.2,0) ;
\draw[very thick] (0.5,1) -- (1.5,1.5) ;
\draw[very thin] (-0.5,1.6) -- (1,2) ;
\draw[very thin] (-0.5,1.6) -- (1.5,1.5) ;
\draw[very thin] (-0.5,1.6) -- (1.2,0) ;
\draw[very thin] (-0.5,1.6) -- (0.5,1) ;
\draw[very thin] (2.5,1.2) -- (1,2) ;
\draw[very thin] (2.5,1.2) -- (1.5,1.5) ;
\draw[very thin] (2.5,1.2) -- (1.2,0) ;
\draw[very thin] (2.5,1.2) -- (0.5,1) ;
\filldraw[black] (1,2) circle (0pt) node[anchor=south] {\footnotesize 1};
\filldraw[black] (1.5,1.5) circle (0pt) node[anchor=north] {\footnotesize\ \ 2};
\filldraw[black] (1.2,0) circle (0pt) node[anchor=north] {\footnotesize 3};
\filldraw[black] (0.5,1) circle (0pt) node[anchor=north] {\footnotesize 4};
\filldraw[black] (-0.5,1.6) circle (0pt) node[anchor=east] {\footnotesize 5};
\filldraw[black] (2.5,1.2) circle (0pt) node[anchor=west] {\footnotesize 6};
\end{tikzpicture}
\end{center}
The first $4$-simplex, $\sigma_1$, is determined by the vertices $(1,2,3,4,5)$, while the second, $\sigma_2$, is determined by the vertices $(1,2,3,4,6)$. They share the common tetrahedron $\tau$ determined by the vertices $(1,2,3,4)$, depicted with thick edges. There are four triangles of $\tau$, shared by both $4$-simplices, namely
$$
(1,2,3)\,, \quad
(1,2,4)\,, \quad
(1,3,4)\,, \quad
(2,3,4)\,.
$$
In addition, $\sigma_1$ contains six more triangles
$$
(5,1,2)\,, \quad (5,1,3)\,, \quad (5,1,4)\,, \quad (5,2,3)\,, \quad (5,2,4)\,, \quad (5,3,4)\,,
$$
while $\sigma_2$ contains its own additional six triangles
$$
(6,1,2)\,, \quad (6,1,3)\,, \quad (6,1,4)\,, \quad (6,2,3)\,, \quad (6,2,4)\,, \quad (6,3,4)\,.
$$
In total, the figure has $E=14$ edges and $F=16$ triangles. For each triangle we write a simplicity constraint equation, giving rise to $16$ equations:
\begin{equation}
\begin{array}{lcl}
|m_{123}| & = & \ds \frac{1}{\gamma l_p^2} A_H(l_{12},l_{13},l_{23})\,, \\
 & \vdots & \\
|m_{634}| & = & \ds \frac{1}{\gamma l_p^2} A_H(l_{63},l_{64},l_{34})\,. \\
\end{array}
\end{equation}
As discussed in the previous case, one can always solve the system of the first ten equations, for the $4$-simplex $\sigma_1$. Denote this solution as
\begin{equation} \label{RnjeZaLeviSimplex}
\begin{array}{ccccccc}
m_{123} & = & \bar{m}_{123}\,, & \quad & l_{12} & = & L_{12}\,, \\
 & \vdots & & & & \vdots & \\
m_{534} & = & \bar{m}_{534}\,, & & l_{54} & = & L_{54}\,. \\
\end{array}
\end{equation}
One could now arguably also solve four of the remaining six equations, for example for triangles $(6,1,2)$, $(6,1,3)$, $(6,1,4)$ and $(6,2,3)$, giving rise to additional four $\bar{m}$'s and four $L$'s. Substituting all previous results into the remaining two equations, one gets
\begin{equation} \label{DvaDiofanta}
|m_{624}| = f(\bar{m},L)\,, \qquad |m_{634}| = g(\bar{m},L) \, .
\end{equation}
Here $f$ and $g$ are two functions implicitly defined by the substitution of $14$ $\bar{m}$'s and $14$ $L$'s into the remaining two equations. However, as discussed previously, there is a priori no guarantee that there will exist a choice for $(\bar{m},L)$ such that $m_{624}$ and $m_{634}$ are integers. But there is a beautiful geometrical argument which proves that this choice indeed exists. Namely, given the generic solution (\ref{RnjeZaLeviSimplex}) for $\sigma_1$, choose the following for $\sigma_2$:
\begin{equation} \label{RnjeZaDesniSimplexLovi}
l_{61} = L_{51}\,, \quad
l_{62} = L_{52}\,, \quad
l_{63} = L_{53}\,, \quad
l_{64} = L_{54}\,.
\end{equation}
In other words, we choose the edges of $\sigma_2$ to be equal to corresponding edges of $\sigma_1$, making the two $4$-simplices identical (up to reflection symmetry). Since the triangle areas of $\sigma_1$ are integers by construction, so will be the triangle areas of $\sigma_2$. In particular, we have
\begin{equation} \label{RnjeZaDesniSimplexMovi}
\begin{array}{ccc}
m_{612} = \bar{m}_{512}\,, & \quad & m_{623} = \bar{m}_{523}\,, \\
m_{613} = \bar{m}_{513}\,, & & m_{624} = \bar{m}_{524}\,, \\
m_{614} = \bar{m}_{514}\,, & & m_{634} = \bar{m}_{534}\,. \\
\end{array}
\end{equation}
In this way, all $16$ simplicity constraint equations are simultaneously satisfied, giving us one explicit solution for the case of two connected $4$-simplices. Numerical search for other solutions of the two equations (\ref{DvaDiofanta}) has proved fruitless, suggesting that this is the only possible generic solution. In special cases (like isosceles $4$-simplices) there may be more solutions, and examples of these will be given in the next section.

Finally, having studied the special cases of one and two $4$-simplices, we now turn to the general case of a triangulation containing arbitrary many simplices. The proof of the existence of a solution to the simplicity constraint (\ref{SimplicitySistemJna}) is simple, and is performed by induction over the number of $4$-simplices. As a base case, note that a single simplex can be labeled with $10$ $\bar{m}$'s and $10$ $L$'s such that the simplicity constraint is satisfied, as already discussed above. To prove the inductive step, first assume that we have found a solution $(\bar{m},L)$ that labels $N$ $4$-simplices. Then, assuming that one of those $4$-simplices features a free boundary tetrahedron, we proceed to attach the $(N+1)$-th $4$-simplex to that boundary tetrahedron. We then specify the labels $(m,l)$ of that additional $4$-simplex to be equal to the corresponding labels of its neighbor, by the recipe discussed in the case of two $4$-simplices above (namely (\ref{RnjeZaDesniSimplexLovi}) and (\ref{RnjeZaDesniSimplexMovi})). The analysis of two $4$-simplices above then guarantees that those labels also satisfy the additional six simplicity constraint equations, since the equations are the same for the two $4$-simplices, and they are satisfied for one of them by induction hypothesis. Therefore, we have constructed an extended set of labels $(\bar{m},L)$ which satisfies the simplicity constraint for $N+1$ $4$-simplices. This completes the proof of the induction step.

At this point a few comments are in order. First, numerical search for solutions distinct from (\ref{RnjeZaDesniSimplexLovi}), (\ref{RnjeZaDesniSimplexMovi}) has failed to find any, suggesting that this solution of the simplicity constraint equations is the only possible one, at least in the generic case. Second, as we can see from the construction, once we choose some particular labeling for the first $4$-simplex, all other $4$-simplices in the triangulation are labeled in the same way, i.e. the triangulation consists of $4$-simplices which are all mutually identical. This has rather nontrivial physical consequences, because the state sum (\ref{SpincubeStateSumaStrong}) reduces to the form
\begin{equation} \label{DrugaKvantizacijaZamaloCDTzaJednuTriang}
Z_{T}^{\rm strong} = \sum_{ \bar{m}_1 \in\celi } \dots \sum_{ \bar{m}_{10} \in\celi } \sum_{\alpha} \; e^{iS_R(L_{\alpha}(\bar{m}))} \,,
\end{equation}
where $\alpha$ counts the number of possible solutions for $L$'s, given a specific set of $\bar{m}$'s, in the system (\ref{SistemJnaZaJedanSimplex}). From the state sum one can see that there are only ten degrees of freedom in the whole spacetime, specified by the identical labeling of all $4$-simplices in $T$. This is unsatisfactory from the physical point of view, because we expect that the theory gives general relativity in the classical limit, and general relativity has two physical degrees of freedom (i.e. two graviton polarizations) per every point in space, which is certainly more than ten in total. This problem is commonly resolved by passing to the formalism of ``second quantization'', i.e. by taking an additional sum over various possible triangulations:
\begin{equation} \label{DrugaKvantizacijaZamaloCDT}
Z^{\rm sq} \equiv \sum_{T\in \cT} Z_{T}^{\rm strong} = \sum_{T\in\cT} \sum_{ \bar{m}_1 \in\celi } \dots \sum_{ \bar{m}_{10} \in\celi } \sum_{\alpha} \; e^{iS_R(L_{\alpha}(\bar{m}))} .
\end{equation}
The set $\cT$ is some nonempty set of inequivalent triangulations $T(\cM_4)$, keeping the topology of the manifold $\cM_4$ the same. One can in principle attempt to discuss ``all'' possible triangulations, but it is often more useful to restrict to a certain class, according to various physical and mathematical criteria. Obviously, one of these criteria is the convergence of the state sum (\ref{DrugaKvantizacijaZamaloCDT}), assuming of course that (\ref{DrugaKvantizacijaZamaloCDTzaJednuTriang}) is finite to begin with.

We will not discuss the convergence issues or the choice of $\cal{T}$ in this paper. We just note that since $\cT$ can be chosen to be suitably large, it can provide an arbitrarily large number of degrees of freedom in the theory, through different configurations of the triangulation. It is however important to emphasize that this is a highly nontrivial step, and there are a priori no guarantees that summing over some set of triangulations will indeed introduce the needed degrees of freedom into the theory. It may still happen that the model contains anomalies and the space of solutions is overconstrained. The presence of anomalies is an important open problem that should be studied in the future.

As a final comment, note that the sums over $\bar{m}$'s and $\alpha$ are labeling all $4$-simplices at the same time, independently of the choice of $T$. Under the implicit assumption of uniform convergence of the sum over triangulations, one is therefore allowed to switch the order of summations in (\ref{DrugaKvantizacijaZamaloCDT}) to obtain
\begin{equation} \label{SQgenStateSuma}
Z^{\rm sq} = \sum_{ \bar{m}_1 \in\celi } \dots \sum_{ \bar{m}_{10} \in\celi } \sum_{\alpha} \left( \sum_{T\in\cT} \; e^{iS_R(L_{\alpha}(\bar{m}))} \right) \,,
\end{equation}
where the term in parentheses becomes very similar to the state sum discussed in the models of CDT approach to quantum gravity. In the next section we turn to the analysis of this relationship.

\section{\label{SecRelationToCDT}Relation to CDT}

In the Causal Dynamical Triangulations approach (see \cite{CDTreview} for a review), one constructs the state sum as
\begin{equation} \label{CDTstateSuma}
Z_{\rm CDT} = \sum_{T\in\cT} \; e^{iS_R(a,b)}\,,
\end{equation}
where the sum goes over a class of triangulations $\cT$ specified by causality requirements, while all $4$-simplices in $T$ are isosceles, i.e. labeled by edge lengths $a$ and $b$, such that one can distinguish foliations of $T$ into spacelike hypersurfaces labeled exclusively by edge lengths $a$, while each two hypersurfaces are connected by edge lengths $b$, which is usually chosen to be timelike rather than spacelike (although this is not mandatory). This means that all tetrahedra within a given hypersurface are equilateral, while all $4$-simplices filling up a slice of spacetime between two hypersurfaces are isosceles. In particular, of all possible isosceles $4$-simplices (for given $a,b$ there are in total $40$ inequivalent ones up to reflections and rotations, see Appendix \ref{ApendiksB}), exactly two are used, depicted by the following two diagrams:
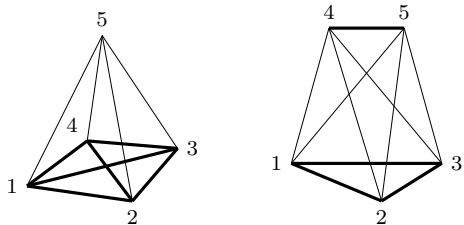
\begin{center}
\begin{tikzpicture}
\draw[very thick] (-1,0) -- (1,0.5) ;
\draw[very thick] (-0.2,0.6) -- (0.4,-0.2) ;
\draw[very thick] (-1,0) -- (-0.2,0.6) ;
\draw[very thick] (-1,0) -- (0.4,-0.2) ;
\draw[very thick] (1,0.5) -- (-0.2,0.6) ;
\draw[very thick] (1,0.5) -- (0.4,-0.2) ;
\draw[very thin] (-1,0) -- (0,2) ;
\draw[very thin] (1,0.5) -- (0,2) ;
\draw[very thin] (-0.2,0.6) -- (0,2) ;
\draw[very thin] (0.4,-0.2) -- (0,2) ;
\filldraw[black] (-1,0) circle (0pt) node[anchor=east] {\footnotesize 1};
\filldraw[black] (0.4,-0.2) circle (0pt) node[anchor=north] {\footnotesize 2};
\filldraw[black] (1,0.5) circle (0pt) node[anchor=west] {\footnotesize 3};
\filldraw[black] (-0.2,0.6) circle (0pt) node[anchor=south east] {\footnotesize 4};
\filldraw[black] (0,2) circle (0pt) node[anchor=south] {\footnotesize 5};
\end{tikzpicture}
\hspace{0.5cm}
\begin{tikzpicture}
\draw[very thick] (-1,0) -- (1,0) ;
\draw[very thick] (-1,0) -- (0.2,-0.5) ;
\draw[very thick] (1,0) -- (0.2,-0.5) ;
\draw[very thick] (-0.5,1.8) -- (0.5,1.8) ;
\draw[very thin] (-1,0) -- (-0.5,1.8) ;
\draw[very thin] (-1,0) -- (0.5,1.8) ;
\draw[very thin] (1,0) -- (-0.5,1.8) ;
\draw[very thin] (1,0) -- (0.5,1.8) ;
\draw[very thin] (0.2,-0.5) -- (-0.5,1.8) ;
\draw[very thin] (0.2,-0.5) -- (0.5,1.8) ;
\filldraw[black] (-1,0) circle (0pt) node[anchor=east] {\footnotesize 1};
\filldraw[black] (0.2,-0.5) circle (0pt) node[anchor=north] {\footnotesize 2};
\filldraw[black] (1,0) circle (0pt) node[anchor=west] {\footnotesize 3};
\filldraw[black] (-0.5,1.8) circle (0pt) node[anchor=south] {\footnotesize 4};
\filldraw[black] (0.5,1.8) circle (0pt) node[anchor=south] {\footnotesize 5};
\end{tikzpicture}
\end{center}
In both diagrams, thick lines are of length $a$, while thin lines are of length $b$. In the notation of \cite{CDTreview} these two $4$-simplices are denoted as $(4,1)$ and $(3,2)$ respectively.

The result of the previous section was the state sum (\ref{SQgenStateSuma}), which is readily comparable to (\ref{CDTstateSuma}). The main difference, however, lies in the fact that for the generic choice of $\bar{m}$'s in (\ref{SQgenStateSuma}) all $4$-simplices in the triangulation must be identical, as dictated by the simplicity constraint. This is in contrast to (\ref{CDTstateSuma}), where two types of $4$-simplices are being used. However, it turns out that this case is also covered by (\ref{SQgenStateSuma}), as we will now show.

As was discussed in the previous section, one can label the whole triangulation by labeling one arbitrary initial $4$-simplex, and then employing the simplicity constraint to fix the labels for all other $4$-simplices. In a generic case, this procedure is unique. However, if the labeling of the initial $4$-simplex is chosen in a special way, there can be more freedom for the choice of other $4$-simplices. As an example, choose the labels of the initial $4$-simplex such that it is isosceles, for example the left one of the two diagrams above. If one assigns length $a$ to thick edges and length $b$ to thin edges, the simplicity constraint for that $4$-simplex reduces to the following two equations:
\begin{equation} \label{SimpConstZaIsoscSimplex}
\gamma l_p^2\, |m_1| = A_H(a,a,a) \equiv \frac{a^2\sqrt{3}}{4}\,, \qquad
\gamma l_p^2\, |m_2| = A_H(a,b,b) \equiv \frac{a}{4} \sqrt{4b^2-a^2}\,.
\end{equation}
This is because only triangles of type $(a,a,a)$ and $(a,b,b)$ appear in the $4$-simplex. Moreover, it is obvious that the simplicity constraint can always be uniquely solved for $a$ and $b$, given any choice of integers $m_1$ and $m_2$:
$$
a = 2l_p \sqrt{\frac{\gamma |m_1|}{\sqrt{3}}}\, , \qquad b = l_p \sqrt{\frac{\gamma |m_1|}{\sqrt{3}}} \sqrt{1+3\left(\frac{m_2}{m_1}\right)^2}\, .
$$

But now note that there is {\it another} $4$-simplex that is also made only of those two triangles, namely the right $4$-simplex in the diagram above. Its simplicity constraint is identical to (\ref{SimpConstZaIsoscSimplex}), and thus already satisfied. Therefore, in this case the $4$-simplices in the triangulation need not be all identical, which leaves more flexibility in the possible labeling of the triangulation. In this case, restricting (\ref{SQgenStateSuma}) to isosceles $4$-simplices of the above type, we obtain
\begin{equation} \label{SpincubeCDTsuma}
Z^{\rm sq} = \sum_{ m_1 \in\celi } \sum_{ m_2 \in\celi } \left( \sum_{T\in\cT} \; e^{iS_R(a,b)} \right) \,,
\end{equation}
where the expression in the parentheses is now precisely the CDT state sum (\ref{CDTstateSuma}). Restricting oneself to this choice of labeling, after ``freezing out'' the values for $m_1$ and $m_2$, the spincube model with the strongly imposed simplicity constraint reduces to the CDT model of quantum gravity.

It is worth noting that there are other isosceles $4$-simplices for which the analogous construction may apply. In particular, there are two more pairs of $4$-simplices which are labeled by only two edge lengths $a,b$ and consist of only two types of triangles (see Appendix \ref{ApendiksB} for details). One can choose those alternative geometries to satisfy the simplicity constraint in a similar way, and obtain a state sum similar to CDT. This is however not equivalent to CDT, because such $4$-simplices do not induce a natural foliation of spacetime into space and time, which is one of the important aspects of CDT.

The presence of the CDT state sum  (\ref{CDTstateSuma}) as one piece of the spincube state sum (\ref{SQgenStateSuma}) essentially means that the spincube model contains the CDT model as a special case. Specifically, if one studies only the isosceles configurations of $4$-simplices within the spincube model, one will recover all the wealth of results that can be obtained within the CDT approach to quantum gravity. In addition the spincube model allows one to study non-CDT-like isosceles configurations mentioned above, as well as non-isosceles configurations, all of which can potentially give rise to novel effects, not present in the CDT approach. Extending the quantum gravity model from the CDT state sum to the spincube state sum is therefore useful, both as a way to study more general spacetime configurations, and as a way to study the relationship between the CDT formalism and the spinfoam formalism, which is a very interesting open problem.

\section{\label{SecTopology}Topological restrictions}

So far we have not discussed the topological and combinatorial properties of the manifold $\cM_4$ and its triangulation $T(\cM_4)$. There are two main questions to be addressed:
\begin{itemize}
\item[(a)] Does the form of the state sum (\ref{SQgenStateSuma}) place any restrictions on the choice of the topology of the manifold $\cM_4$?
\item[(b)] Does the choice of the topology of $\cM_4$ place any restrictions on the form of (\ref{SQgenStateSuma})?
\end{itemize}

Before we even begin discussing these questions, two remarks are necessary. First, it should be noted that we are interested in the realistic choice of four spacetime dimensions, which means that we are discussing $4$-dimensional topological manifolds. The full classification of $4$-dimensional manifolds is known to be an undecidable problem in the sense of G\"odel's first incompleteness theorem \cite{Markov1958,Poonen2014}, and even among simply-connected $4$-manifolds there are those which do not admit a triangulation to begin with, like the $E_8$ manifold \cite{Freedman1982,Scorpan2005}. It is thus clear that any conclusive analysis of the above two questions is hopeless. This section is therefore devoted to discussing some examples and reformulations of the questions, without giving any general answers.

Second, one should keep in mind that the topology of $\cM_4$ is kept fixed when summing over different triangulations in (\ref{SQgenStateSuma}). Given one initial triangulation $T(\cM_4)$, one can construct other, different triangulations of the manifold with the same topology using, for example, Pachner moves. Some suitable subset of these triangulations will define the domain $\cal{T}$ for the sum over triangulations. Additionally summing over different topologies of $\cM_4$ would represent the formalism of third quantization,
$$
Z^{\rm tq} = \sum_{\text{topologies}} Z^{\rm sq},
$$
and is out of the scope of this paper.

After these introductory remarks, we can concentrate on what can be said regarding the questions (a) and (b) above. To this end, it is instructive to rewrite the state sum (\ref{SQgenStateSuma}) as
$$
Z^{\rm sq} = Z_{10} + \dots + Z_1,
$$
where
$$
\begin{array}{ccl}
Z_{10} & = & \ds \sum_{ \substack{ \bar{m}_1,\dots,\bar{m}_{10} \\ \bar{m}_1 \neq \dots \neq \bar{m}_{10} } } \sum_{\alpha} \left( \sum_{T\in\cT} \; e^{iS_R(L_{\alpha}(\bar{m}))} \right)\,,  \\
Z_9 & = & \ds \sum_{ \substack{ \bar{m}_1,\dots,\bar{m}_9 \\ \bar{m}_1 \neq \dots \neq \bar{m}_9 } } \sum_{\alpha} \left( \sum_{T\in\cT} \; e^{iS_R(L_{\alpha}(\bar{m}))} \right) \,, \\
 & \vdots & \\
Z_2 & = & \ds \sum_{ \substack{ \bar{m}_1,\bar{m}_2 \\ \bar{m}_1 \neq \bar{m}_2 } } \sum_{\alpha} \left( \sum_{T\in\cT} \; e^{iS_R(L_{\alpha}(\bar{m}))} \right) \,,  \\
Z_1 & = & \ds \sum_{ \bar{m}_1 } \sum_{\alpha} \left( \sum_{T\in\cT} \; e^{iS_R(L_{\alpha}(\bar{m}))} \right) \,.  \\
\end{array}
$$
Here we have rearranged the summation over the ten $\bar{m}$ integers such that we can explicitly distinguish the pieces of the state sum where two, three, etc. of them are equal. The $Z_{10}$ term corresponds to the generic situation when all $\bar{m}$'s are mutually different. The $Z_1$ term corresponds to the case where all of $\bar{m}$'s are equal. The $Z_2$ term contains the generalized CDT state sum (\ref{SpincubeCDTsuma}). In relation to questions (a) and (b) above, it will turn out that the $Z_{10}$ term will be most useful to discuss (b), while the $Z_1$ term is the most useful to discuss (a).

The second ingredient we need is the $1$-complex dual to the triangulation. It is constructed as follows: to each $4$-simplex in the triangulation one assigns a vertex, and to each tetrahedron common to two $4$-simplices one assigns a link connecting two vertices. If the tetrahedron is on the boundary, then it belongs to only one $4$-simplex, and the corresponding link is open-ended. Thus the $1$-complex dual to the triangulation is some $5$-valent graph with links encoding the adjacency relation among $4$-simplices, while open-ended links encode boundary tetrahedra of the triangulation.

Looking first at the generic case of $Z_{10}$, it was emphasized in section \ref{SecSolutionToTheSimplicityConstraint} that all $4$-simplices must be identical, while having $5$ different irregular tetrahedra on their boundaries. The $4$-simplices are then glued to each other along the identical tetrahedra, keeping the simplicity constraint satisfied. However, there exist a possibility that, after a certain number of gluings, two distant parts of the triangulation are supposed to ``meet'' due to topological structure, and at the meeting point they may fail to have compatible tetrahedra for consistent gluing. In this way topology may invalidate the generic choice of $\bar{m}$'s in $Z_{10}$, and thus place a restriction the form of the state sum $Z^{\rm sq}$ (see question (b)).

This problem is actually equivalent to the problem of consistently coloring a $1$-complex dual to the triangulation such that each link is colored by a number from the set $\{1,\dots,5\}$, and such that at every vertex no two colors are repeated. We do not have a general answer to this question. Indeed, there may be $1$-complex graphs for which such a coloring is possible, and graphs for which it is impossible. One obvious example of the former is the construction where we start from one vertex, connect five new vertices to its five links, and then keep connecting new vertices to new free links. But we do this in such a way that we never mutually connect two already existing free links, i.e. we never make a loop. The $1$-complex graph constructed in such a way can always be colored in the required way --- one starts by arbitrarily coloring the links of the initial vertex, and expands the coloring from there, keeping the coloring of links entering into each new vertex consistent with the previous choices. For the opposite example, we do not have any obvious examples, but it is certainly possible that for some graphs consistent coloring may be impossible.

The above analysis suggests that the topology of the manifold might exclude the generic piece $Z_{10}$ from the total state sum, but it might also accommodate it without problems. Regarding the question (b) above, this is the best possible answer one can give at this stage --- depending on the actual choice of the topology of $\cM_4$, there might or might not be some restrictions on the state sum.

On the other extreme, let us look at $Z_1$. This piece of the state sum contains equilateral $4$-simplices. In an equilateral $4$-simplex, there is only one type of tetrahedron available, and any two $4$-simplices can be glued along any of their tetrahedra. In the language of the dual $1$-complex, this means that we are supposed to color the links of the graph with a single color (vertices are now such that all ``five'' colors must be mutually identical). It is pretty obvious that every graph can be colored with a single color, without restrictions. This means that a triangulation corresponding to arbitrary topology can always be labeled with equilateral $4$-simplices.

The problem with this, however, is that equilateral $4$-simplices can only be embedded into a spacetime of Euclidean signature, while we are interested in the realistic Lorentzian case. So the equilateral $4$-simplices actually have to be excluded from the state sum on the grounds of desired spacetime signature. So the only available $4$-simplices in $Z_1$ are isosceles, with at least three different types of distinct tetrahedra. In the language of the dual $1$-complex, this means that we are supposed to label the links of the graph with three distinct colors $\{ 1,2,3 \}$, along with certain combinatorial rules for the vertices. Whether or not this is possible for a graph corresponding to the triangulation of an arbitrary topology, is again a problem for which we have no answer. Like in the generic case, one can certainly construct graphs such that this coloring is possible. In particular, every graph that can be consistently colored with five different colors (corresponding to the $Z_{10}$ piece of the state sum), can also be colored using only three distinct colors (simply by identifying the fourth and fifth color with one of the previous three). For the opposite case we again do not have an explicit example, but it might exist.

The analysis of the $Z_1$ case then suggests an answer to the question (a), albeit an inconclusive one --- the state sum may place restrictions on the possible topologies of $\cM_4$, but also maybe does not, depending on the existence of the dual $1$-complex which cannot be colored with three distinct colors.

Finally, all remaining pieces of the state sum, $Z_9$ to $Z_2$ are ``in between'' the two extreme cases $Z_{10}$ and $Z_1$, and add no new insight. As one goes from $Z_1$ to $Z_{10}$, each piece $Z_k$ for higher $k$ places more stringent rules for coloring the graph, ranging from three colors with repetitions at vertices to five colors without repetitions. In this sense their analysis does not add any novel information about possible topologies that is not already present in cases $Z_1$ and $Z_{10}$.

The inconclusiveness of the answers to questions (a) and (b) given by the qualitative analysis of this section might seem underwhelming and disappointing. One should therefore keep in mind that the undecidability of the classification of topological $4$-manifolds severely limits what one can say about any general question involving the interaction between the structure of the state sum and topology, in the following way. Regarding the question (b), topology of $\cM_4$ places restrictions on the state sum if no triangulation of $\cM_4$ can accommodate the $Z_{10}$ piece of the state sum. If we denote as $S$ the set of all triangulations $T$ that have some chosen fixed topology of $\cM_4$,
$$
S = \{  T \; | \; T \text{ is homeomorphic to } \cM_4 \},
$$
then one can rewrite the statement that topology places restrictions on the state sum as follows:
$$
\forall T \;\; T \in S \Rightarrow \left(
\begin{array}{l}
\text{dual graph of } T \text{ cannot be colored} \\
\text{in a way compatible with } Z_{10} \\
\end{array}
\right)\, .
$$
But this statement is impossible to prove, since it is undecidable whether any given triangulation has the topology of $\cM_4$, so the set of triangulations with appropriate topology, $S$, cannot be effectively quantified over. In other words, the antecedent in the implication is undecidable, which makes the whole sentence undecidable.

Regarding the question (a), the state sum places restrictions on the topology of $\cM_4$ if there is no triangulation of $\cM_4$ that can accommodate the $Z_1$ piece of the state sum:
$$
\forall T \;\; T \in S \Rightarrow \left(
\begin{array}{l}
\text{dual graph of } T \text{ cannot be colored} \\
\text{in a way compatible with } Z_1 \\
\end{array}
\right)\, .
$$
Again, this is impossible to prove, for the same reason as before --- it is undecidable whether or not any given triangulation has the topology of $\cM_4$, so the set of candidate triangulations cannot be quantified over.

Nevertheless, the analysis presented above gives us some practical tools and techniques to study particular cases. For example, given some particular triangulation, one can always construct its dual $1$-complex, and determine whether or not it can be colored according to the rules given above. This question is decidable for any finite graph, since there are finitely many possible choices for the color of each link. One can simply try out all possible combinations of colors for all available links in the graph, and the algorithm is guaranteed to halt after a finite number of steps, since there are finite number of possibilities to test.

\section{\label{SecConclusions}Conclusions}

In this paper we have studied the possibility of strongly imposing the simplicity constraint in the spincube model of quantum gravity. After a short overview of the model, we have introduced two ways to impose the simplicity constraint, weakly and strongly. Weak constraint holds only in the classical limit of the theory, and naturally leads to the Regge model of quantum gravity. However, imposing the constraint strongly, such that it holds at the quantum level, leads to a nontrivial issue of consistency which depends on solving a complicated system of Diophantine equations. As the main result of the paper, it was shown that this Diophantine equations indeed do have solutions, and two classes of solutions were constructed. The first class represents the case when all $4$-simplices in the triangulation are identical (up to reflection and rotation symmetries). The second class represents the case of isosceles $4$-simplices, whose edge lengths take only two different values. Moreover, one can choose the labeling of those isosceles $4$-simplices to be such that the theory is equivalent to the state sum of the CDT approach to quantum gravity. Finally, we have discussed the topological properties of the state sum, where the conclusions are very limited by the undecidability of the classification of topological $4$-dimensional manifolds.

We close with a few remarks. Note that the spincube model state sums (\ref{SQgenStateSuma}) and (\ref{SpincubeCDTsuma}) are more general than the CDT state sum (\ref{CDTstateSuma}), in the sense that one sums over all possible choices for the triangle labels $m_1$ and $m_2$, which are kept constant in the CDT model. The reason for this is that we have extended the spincube model to the formalism of second quantization --- summing both over the field variables and the triangulations, rather than summing over only one of those. As was discussed, the second quantization formalism is necessary if one strongly imposes the simplicity constraint, since otherwise the model has too few degrees of freedom. Of course, passing on to the formalism of second quantization does not automatically guarantee the introduction of the correct number of degrees of freedom, since the theory may contain anomalies and the space of solutions may be overconstrained. This is common for both the spincube model and the CDT approach, and is an open problem that should be studied in the future.

On the other hand, summing over both the field variables and the triangulations provides us with a better understanding of the theory. Namely, one can resolve the long-standing problem of the relationship between the CDT approach and various covariant LQG approaches (spinfoam and spincube models), within the same unifying framework. Also, the formalism includes other types of solutions to the simplicity constraint and therefore other specific models inequivalent to CDT. In this sense, one can study those alternative models and compare them to the CDT approach. This opens many novel avenues for further research.

Finally, one can study some nontrivial issues related to the classical limit. Namely, in CDT the classical limit is defined statistically, by averaging over many possible triangulations with a given (large) number of $4$-simplices. Depending on the parameters in the action, one discovers three different phases in the theory, one of which can be related to the classical general relativity. In this setup the edge lengths of the $4$-simplices are mutually equal and are kept fixed. On the other hand, in the spinfoam and spincube models, the classical limit is obtained via the asymptotic limit in which edge lengths, areas and volumes become very large, while the triangulation itself is kept fixed. It is an open question whether these two limiting procedures are equivalent or not, what is the meaning of each, and which procedure is more appropriate for the study of the classical theory emerging from the model. The second quantization formalism of the spincube model discussed in this paper provides one with the tools to study both limits within the same model. In particular, it would be interesting to reproduce the phase diagram of CDT geometries for a model consisting of irregular $4$-simplices, where more than three edges (up to all ten) have different values. One could study the properties of resulting phases, and how they relate to the CDT phases as one varies the values of the various edge length parameters. Also, one could discuss whether in the limit of large edge lengths one always ends up in the same CDT phase or not, and why. All these questions are interesting to study, and will be addressed in future work.

\begin{acknowledgments}
The author would like to thank Aleksandar Mikovi\'c, Nikola Paunkovi\'c and the three anonymous Referees for helpful comments and suggestions. This work was supported by the FCT grant SFRH/BPD/46376/2008, the FCT project PEst-OE/MAT/UI0208/2013, and partially by the project ON171031 of the Ministry of Education, Science and Technological Development, Serbia.
\end{acknowledgments}

\appendix

\section{\label{ApendiksA}State Sum Models}

A general state sum model in $D$ dimensions is constructed as follows. Let $T(\cM_D)$ be a triangulation of the manifold $\cM_D$. Triangulation $T$ contains vertices $v$, edges $\epsilon$, triangles $\Delta$, tetrahedra $\tau$, and so on up to $D$-simplices $\sigma$. Each of these objects is labeled with ``colors'' $\phi$, a set of quantities that describe the fundamental variables of the model. In addition, each type of objects in $T$ is assigned an ``amplitude'' $\cA$, which enters the description of the dynamics of variables $\phi$. One then writes the state sum $Z$ as
\begin{equation} \label{GeneralStateSumModel}
Z = \sum_{\{ \phi\} } \; \prod_{v\in T} \cA_v(\phi) \prod_{\epsilon\in T} \cA_{\epsilon}(\phi) \dots \prod_{\sigma\in T} \cA_{\sigma}(\phi) \,.
\end{equation}
The sum goes over the full domain of each color $\phi$ living on each vertex, edge, triangle, tetrahedron or a higher order simplex.

The above expression should be understood as a rigorous definition of a path integral
\begin{equation} \label{GeneralContinuumPathIntegral}
Z = \int \cD\phi \; \exp \left( i S[\phi] \vphantom{\sum} \right)\, ,
\end{equation}
where the $D+1$ amplitude functions $\cA_v$, $\cA_{\epsilon}$, $\cA_{\Delta}$, $\dots$, $\cA_{\sigma}$ partly enter into the definition of the path integral measure $\cD\phi$ and partly into the definition of the action $S[\phi]$. Roughly speaking, specifying the set of variables $\phi$ and their domain is equivalent to the specification of the fields present in the model, while specifying the amplitudes $\cA$ is equivalent to choosing their dynamics.

The equation (\ref{GeneralStateSumModel}) represents the discretization of (\ref{GeneralContinuumPathIntegral}) in the sense of Feynman. In particular, if one constructs a state sum model over a one-dimensional manifold $\cM_1=\realni$ (called the time axis) and chooses amplitudes $\cA_v$ and $\cA_{\epsilon}$ conveniently, equation (\ref{GeneralStateSumModel}) reduces precisely to the Feynman's textbook ``definition by discretization'' of the configuration-space path integral in quantum mechanics. Generalization of the same idea to multiple-dimensional manifold $\cM_D$ leads to (\ref{GeneralStateSumModel}) in the general case.

\section{\label{ApendiksB}Isosceles $4$-simplices}

We define a given $4$-simplex to be called ``isosceles'' iff all its edges have length $a$ or $b$, where by convention $a,b\in\realni^+$. All triangles in such a simplex are either isosceles or equilateral. Also by convention, we exclude fully equilateral $4$-simplices, by requiring that edges of both lengths $a$ and $b$ must be present in the $4$-simplex.

It is an interesting question to determine all possible inequivalent isosceles $4$-simplices. There are in total $20$ different ones, up to the $a\leftrightarrow b$ exchange symmetry and permutations of the vertices. If we label the vertices of a $4$-simplex as per the diagram below,
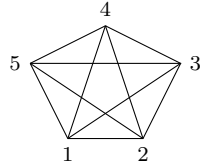
\begin{center}
\begin{tikzpicture}
\draw (-0.5,0) -- (0.5,0) ;
\draw (-0.5,0) -- (1,1) ;
\draw (-0.5,0) -- (-1,1) ;
\draw (-0.5,0) -- (0,1.5) ;
\draw (-1,1) -- (1,1) ;
\draw (-1,1) -- (0,1.5) ;
\draw (-1,1) -- (0.5,0) ;
\draw (1,1) -- (0,1.5) ;
\draw (1,1) -- (0.5,0) ;
\draw (0.5,0) -- (0,1.5) ;
\filldraw[black] (-0.5,0) circle (0pt) node[anchor=north] {\footnotesize 1};
\filldraw[black] (0.5,0) circle (0pt) node[anchor=north] {\footnotesize 2};
\filldraw[black] (1,1) circle (0pt) node[anchor=west] {\footnotesize 3};
\filldraw[black] (0,1.5) circle (0pt) node[anchor=south] {\footnotesize 4};
\filldraw[black] (-1,1) circle (0pt) node[anchor=east] {\footnotesize 5};
\end{tikzpicture}
\end{center}
and if we order the ten edge lengths into a $10$-tuple as $(l_{12},l_{13},l_{14},l_{15},l_{23},l_{24},l_{25},l_{34},l_{35},l_{45})$, the distinct labelings for the $20$ isosceles $4$-simplices are given as:
$$
\begin{array}{rcc}
 1: & & (a,a,a,a,a,a,a,a,a,b)\,, \\
 2: & & (a,a,a,a,a,a,a,a,b,b)\,, \\
 3: & & (a,a,a,a,a,a,b,b,a,a)\,, \\
 4: & & (a,a,a,a,a,a,b,a,b,b)\,, \\
 5: & & (a,a,a,b,a,a,b,b,a,a)\,, \\
 6: & & (a,a,a,a,a,a,b,b,a,b)\,, \\
 7: & & (a,a,a,a,a,a,a,b,b,b)\,, \\
 8: & & (b,b,a,a,a,b,a,a,a,a)\,, \\
 9: & & (a,a,a,b,a,a,b,a,b,b)\,, \\
10: & & (a,a,a,b,a,a,b,b,a,b)\,, \\
\end{array} \qquad
\begin{array}{rcc}
11: & & (a,a,a,b,a,a,b,b,a,b)\,, \\
12: & & (a,a,a,b,b,b,a,b,a,a)\,, \\
13: & & (a,a,a,b,a,b,a,b,b,a)\,, \\
14: & & (a,a,a,a,a,a,b,b,b,b)\,, \\
15: & & (a,a,b,b,b,a,b,b,a,a)\,, \\
16: & & (a,a,a,b,a,b,b,b,b,a)\,, \\
17: & & (a,a,a,b,a,a,b,b,b,b)\,, \\
18: & & (a,a,a,b,a,b,a,b,b,b)\,, \\
19: & & (a,a,a,a,a,b,b,b,b,b)\,, \\
20: & & (a,a,a,b,b,b,a,b,a,b)\,. \\
\end{array}
$$
The above table has been computer-generated by brute force counting of all possibilities. If we mark all $a$-edges with thin lines and all $b$-edges with thick lines, these $4$-simplices can be drawn as follows, respectively:
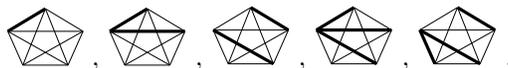
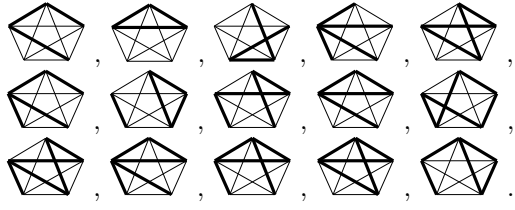
\begin{center}
\begin{tikzpicture}[yscale=1.5]
\draw[very thin] (-0.3,0) -- (0.3,0) ;
\draw[very thin] (-0.3,0) -- (0.5,0.3) ;
\draw[very thin] (-0.3,0) -- (0,0.5) ;
\draw[very thin] (-0.3,0) -- (-0.5,0.3) ;
\draw[very thin] (0.5,0.3) -- (0.3,0) ;
\draw[very thin] (0.3,0) -- (0,0.5) ;
\draw[very thin] (-0.5,0.3) -- (0.3,0) ;
\draw[very thin] (0.5,0.3) -- (0,0.5) ;
\draw[very thin] (-0.5,0.3) -- (0.5,0.3) ;
\draw[very thick] (-0.5,0.3) -- (0,0.5) ;
\end{tikzpicture}
,\ 
\begin{tikzpicture}[yscale=1.5]
\draw[very thin] (-0.3,0) -- (0.3,0) ;
\draw[very thin] (-0.3,0) -- (0.5,0.3) ;
\draw[very thin] (-0.3,0) -- (0,0.5) ;
\draw[very thin] (-0.3,0) -- (-0.5,0.3) ;
\draw[very thin] (0.5,0.3) -- (0.3,0) ;
\draw[very thin] (0.3,0) -- (0,0.5) ;
\draw[very thin] (-0.5,0.3) -- (0.3,0) ;
\draw[very thin] (0.5,0.3) -- (0,0.5) ;
\draw[very thick] (-0.5,0.3) -- (0.5,0.3) ;
\draw[very thick] (-0.5,0.3) -- (0,0.5) ;
\end{tikzpicture}
,\ 
\begin{tikzpicture}[yscale=1.5]
\draw[very thin] (-0.3,0) -- (0.3,0) ;
\draw[very thin] (-0.3,0) -- (0.5,0.3) ;
\draw[very thin] (-0.3,0) -- (0,0.5) ;
\draw[very thin] (-0.3,0) -- (-0.5,0.3) ;
\draw[very thin] (0.5,0.3) -- (0.3,0) ;
\draw[very thin] (0.3,0) -- (0,0.5) ;
\draw[very thick] (-0.5,0.3) -- (0.3,0) ;
\draw[very thick] (0.5,0.3) -- (0,0.5) ;
\draw[very thin] (-0.5,0.3) -- (0.5,0.3) ;
\draw[very thin] (-0.5,0.3) -- (0,0.5) ;
\end{tikzpicture}
,\ 
\begin{tikzpicture}[yscale=1.5]
\draw[very thin] (-0.3,0) -- (0.3,0) ;
\draw[very thin] (-0.3,0) -- (0.5,0.3) ;
\draw[very thin] (-0.3,0) -- (0,0.5) ;
\draw[very thin] (-0.3,0) -- (-0.5,0.3) ;
\draw[very thin] (0.5,0.3) -- (0.3,0) ;
\draw[very thin] (0.3,0) -- (0,0.5) ;
\draw[very thick] (-0.5,0.3) -- (0.3,0) ;
\draw[very thin] (0.5,0.3) -- (0,0.5) ;
\draw[very thick] (-0.5,0.3) -- (0.5,0.3) ;
\draw[very thick] (-0.5,0.3) -- (0,0.5) ;
\end{tikzpicture}
,\ 
\begin{tikzpicture}[yscale=1.5]
\draw[very thin] (-0.3,0) -- (0.3,0) ;
\draw[very thin] (-0.3,0) -- (0.5,0.3) ;
\draw[very thin] (-0.3,0) -- (0,0.5) ;
\draw[very thick] (-0.3,0) -- (-0.5,0.3) ;
\draw[very thin] (0.5,0.3) -- (0.3,0) ;
\draw[very thin] (0.3,0) -- (0,0.5) ;
\draw[very thick] (-0.5,0.3) -- (0.3,0) ;
\draw[very thick] (0.5,0.3) -- (0,0.5) ;
\draw[very thin] (-0.5,0.3) -- (0.5,0.3) ;
\draw[very thin] (-0.5,0.3) -- (0,0.5) ;
\end{tikzpicture}
, \\
\begin{tikzpicture}[yscale=1.5]
\draw[very thin] (-0.3,0) -- (0.3,0) ;
\draw[very thin] (-0.3,0) -- (0.5,0.3) ;
\draw[very thin] (-0.3,0) -- (0,0.5) ;
\draw[very thin] (-0.3,0) -- (-0.5,0.3) ;
\draw[very thin] (0.5,0.3) -- (0.3,0) ;
\draw[very thin] (0.3,0) -- (0,0.5) ;
\draw[very thick] (-0.5,0.3) -- (0.3,0) ;
\draw[very thick] (0.5,0.3) -- (0,0.5) ;
\draw[very thin] (-0.5,0.3) -- (0.5,0.3) ;
\draw[very thick] (-0.5,0.3) -- (0,0.5) ;
\end{tikzpicture}
,\ 
\begin{tikzpicture}[yscale=1.5]
\draw[very thin] (-0.3,0) -- (0.3,0) ;
\draw[very thin] (-0.3,0) -- (0.5,0.3) ;
\draw[very thin] (-0.3,0) -- (0,0.5) ;
\draw[very thin] (-0.3,0) -- (-0.5,0.3) ;
\draw[very thin] (0.5,0.3) -- (0.3,0) ;
\draw[very thin] (0.3,0) -- (0,0.5) ;
\draw[very thin] (-0.5,0.3) -- (0.3,0) ;
\draw[very thick] (0.5,0.3) -- (0,0.5) ;
\draw[very thick] (-0.5,0.3) -- (0.5,0.3) ;
\draw[very thick] (-0.5,0.3) -- (0,0.5) ;
\end{tikzpicture}
,\ 
\begin{tikzpicture}[yscale=1.5]
\draw[very thick] (-0.3,0) -- (0.3,0) ;
\draw[very thick] (-0.3,0) -- (0.5,0.3) ;
\draw[very thin] (-0.3,0) -- (0,0.5) ;
\draw[very thin] (-0.3,0) -- (-0.5,0.3) ;
\draw[very thin] (0.5,0.3) -- (0.3,0) ;
\draw[very thick] (0.3,0) -- (0,0.5) ;
\draw[very thin] (-0.5,0.3) -- (0.3,0) ;
\draw[very thin] (0.5,0.3) -- (0,0.5) ;
\draw[very thin] (-0.5,0.3) -- (0.5,0.3) ;
\draw[very thin] (-0.5,0.3) -- (0,0.5) ;
\end{tikzpicture}
,\ 
\begin{tikzpicture}[yscale=1.5]
\draw[very thin] (-0.3,0) -- (0.3,0) ;
\draw[very thin] (-0.3,0) -- (0.5,0.3) ;
\draw[very thin] (-0.3,0) -- (0,0.5) ;
\draw[very thick] (-0.3,0) -- (-0.5,0.3) ;
\draw[very thin] (0.5,0.3) -- (0.3,0) ;
\draw[very thin] (0.3,0) -- (0,0.5) ;
\draw[very thick] (-0.5,0.3) -- (0.3,0) ;
\draw[very thin] (0.5,0.3) -- (0,0.5) ;
\draw[very thick] (-0.5,0.3) -- (0.5,0.3) ;
\draw[very thick] (-0.5,0.3) -- (0,0.5) ;
\end{tikzpicture}
,\ 
\begin{tikzpicture}[yscale=1.5]
\draw[very thin] (-0.3,0) -- (0.3,0) ;
\draw[very thin] (-0.3,0) -- (0.5,0.3) ;
\draw[very thin] (-0.3,0) -- (0,0.5) ;
\draw[very thin] (-0.3,0) -- (-0.5,0.3) ;
\draw[very thin] (0.5,0.3) -- (0.3,0) ;
\draw[very thick] (0.3,0) -- (0,0.5) ;
\draw[very thick] (-0.5,0.3) -- (0.3,0) ;
\draw[very thick] (0.5,0.3) -- (0,0.5) ;
\draw[very thick] (-0.5,0.3) -- (0.5,0.3) ;
\draw[very thin] (-0.5,0.3) -- (0,0.5) ;
\end{tikzpicture}
, \\
\begin{tikzpicture}[yscale=1.5]
\draw[very thin] (-0.3,0) -- (0.3,0) ;
\draw[very thin] (-0.3,0) -- (0.5,0.3) ;
\draw[very thin] (-0.3,0) -- (0,0.5) ;
\draw[very thick] (-0.3,0) -- (-0.5,0.3) ;
\draw[very thin] (0.5,0.3) -- (0.3,0) ;
\draw[very thin] (0.3,0) -- (0,0.5) ;
\draw[very thick] (-0.5,0.3) -- (0.3,0) ;
\draw[very thick] (0.5,0.3) -- (0,0.5) ;
\draw[very thin] (-0.5,0.3) -- (0.5,0.3) ;
\draw[very thick] (-0.5,0.3) -- (0,0.5) ;
\end{tikzpicture}
,\ 
\begin{tikzpicture}[yscale=1.5]
\draw[very thin] (-0.3,0) -- (0.3,0) ;
\draw[very thin] (-0.3,0) -- (0.5,0.3) ;
\draw[very thin] (-0.3,0) -- (0,0.5) ;
\draw[very thick] (-0.3,0) -- (-0.5,0.3) ;
\draw[very thick] (0.5,0.3) -- (0.3,0) ;
\draw[very thick] (0.3,0) -- (0,0.5) ;
\draw[very thin] (-0.5,0.3) -- (0.3,0) ;
\draw[very thick] (0.5,0.3) -- (0,0.5) ;
\draw[very thin] (-0.5,0.3) -- (0.5,0.3) ;
\draw[very thin] (-0.5,0.3) -- (0,0.5) ;
\end{tikzpicture}
,\ 
\begin{tikzpicture}[yscale=1.5]
\draw[very thin] (-0.3,0) -- (0.3,0) ;
\draw[very thin] (-0.3,0) -- (0.5,0.3) ;
\draw[very thin] (-0.3,0) -- (0,0.5) ;
\draw[very thick] (-0.3,0) -- (-0.5,0.3) ;
\draw[very thin] (0.5,0.3) -- (0.3,0) ;
\draw[very thick] (0.3,0) -- (0,0.5) ;
\draw[very thin] (-0.5,0.3) -- (0.3,0) ;
\draw[very thick] (0.5,0.3) -- (0,0.5) ;
\draw[very thick] (-0.5,0.3) -- (0.5,0.3) ;
\draw[very thin] (-0.5,0.3) -- (0,0.5) ;
\end{tikzpicture}
,\ 
\begin{tikzpicture}[yscale=1.5]
\draw[very thin] (-0.3,0) -- (0.3,0) ;
\draw[very thin] (-0.3,0) -- (0.5,0.3) ;
\draw[very thin] (-0.3,0) -- (0,0.5) ;
\draw[very thin] (-0.3,0) -- (-0.5,0.3) ;
\draw[very thin] (0.5,0.3) -- (0.3,0) ;
\draw[very thin] (0.3,0) -- (0,0.5) ;
\draw[very thick] (-0.5,0.3) -- (0.3,0) ;
\draw[very thick] (0.5,0.3) -- (0,0.5) ;
\draw[very thick] (-0.5,0.3) -- (0.5,0.3) ;
\draw[very thick] (-0.5,0.3) -- (0,0.5) ;
\end{tikzpicture}
,\ 
\begin{tikzpicture}[yscale=1.5]
\draw[very thin] (-0.3,0) -- (0.3,0) ;
\draw[very thin] (-0.3,0) -- (0.5,0.3) ;
\draw[very thick] (-0.3,0) -- (0,0.5) ;
\draw[very thick] (-0.3,0) -- (-0.5,0.3) ;
\draw[very thick] (0.5,0.3) -- (0.3,0) ;
\draw[very thin] (0.3,0) -- (0,0.5) ;
\draw[very thick] (-0.5,0.3) -- (0.3,0) ;
\draw[very thick] (0.5,0.3) -- (0,0.5) ;
\draw[very thin] (-0.5,0.3) -- (0.5,0.3) ;
\draw[very thin] (-0.5,0.3) -- (0,0.5) ;
\end{tikzpicture}
, \\
\begin{tikzpicture}[yscale=1.5]
\draw[very thin] (-0.3,0) -- (0.3,0) ;
\draw[very thin] (-0.3,0) -- (0.5,0.3) ;
\draw[very thin] (-0.3,0) -- (0,0.5) ;
\draw[very thick] (-0.3,0) -- (-0.5,0.3) ;
\draw[very thin] (0.5,0.3) -- (0.3,0) ;
\draw[very thick] (0.3,0) -- (0,0.5) ;
\draw[very thick] (-0.5,0.3) -- (0.3,0) ;
\draw[very thick] (0.5,0.3) -- (0,0.5) ;
\draw[very thick] (-0.5,0.3) -- (0.5,0.3) ;
\draw[very thin] (-0.5,0.3) -- (0,0.5) ;
\end{tikzpicture}
,\ 
\begin{tikzpicture}[yscale=1.5]
\draw[very thin] (-0.3,0) -- (0.3,0) ;
\draw[very thin] (-0.3,0) -- (0.5,0.3) ;
\draw[very thin] (-0.3,0) -- (0,0.5) ;
\draw[very thick] (-0.3,0) -- (-0.5,0.3) ;
\draw[very thin] (0.5,0.3) -- (0.3,0) ;
\draw[very thin] (0.3,0) -- (0,0.5) ;
\draw[very thick] (-0.5,0.3) -- (0.3,0) ;
\draw[very thick] (0.5,0.3) -- (0,0.5) ;
\draw[very thick] (-0.5,0.3) -- (0.5,0.3) ;
\draw[very thick] (-0.5,0.3) -- (0,0.5) ;
\end{tikzpicture}
,\ 
\begin{tikzpicture}[yscale=1.5]
\draw[very thin] (-0.3,0) -- (0.3,0) ;
\draw[very thin] (-0.3,0) -- (0.5,0.3) ;
\draw[very thin] (-0.3,0) -- (0,0.5) ;
\draw[very thick] (-0.3,0) -- (-0.5,0.3) ;
\draw[very thin] (0.5,0.3) -- (0.3,0) ;
\draw[very thick] (0.3,0) -- (0,0.5) ;
\draw[very thin] (-0.5,0.3) -- (0.3,0) ;
\draw[very thick] (0.5,0.3) -- (0,0.5) ;
\draw[very thick] (-0.5,0.3) -- (0.5,0.3) ;
\draw[very thick] (-0.5,0.3) -- (0,0.5) ;
\end{tikzpicture}
,\ 
\begin{tikzpicture}[yscale=1.5]
\draw[very thin] (-0.3,0) -- (0.3,0) ;
\draw[very thin] (-0.3,0) -- (0.5,0.3) ;
\draw[very thin] (-0.3,0) -- (0,0.5) ;
\draw[very thin] (-0.3,0) -- (-0.5,0.3) ;
\draw[very thin] (0.5,0.3) -- (0.3,0) ;
\draw[very thick] (0.3,0) -- (0,0.5) ;
\draw[very thick] (-0.5,0.3) -- (0.3,0) ;
\draw[very thick] (0.5,0.3) -- (0,0.5) ;
\draw[very thick] (-0.5,0.3) -- (0.5,0.3) ;
\draw[very thick] (-0.5,0.3) -- (0,0.5) ;
\end{tikzpicture}
,\ 
\begin{tikzpicture}[yscale=1.5]
\draw[very thin] (-0.3,0) -- (0.3,0) ;
\draw[very thin] (-0.3,0) -- (0.5,0.3) ;
\draw[very thin] (-0.3,0) -- (0,0.5) ;
\draw[very thick] (-0.3,0) -- (-0.5,0.3) ;
\draw[very thick] (0.5,0.3) -- (0.3,0) ;
\draw[very thick] (0.3,0) -- (0,0.5) ;
\draw[very thin] (-0.5,0.3) -- (0.3,0) ;
\draw[very thick] (0.5,0.3) -- (0,0.5) ;
\draw[very thin] (-0.5,0.3) -- (0.5,0.3) ;
\draw[very thick] (-0.5,0.3) -- (0,0.5) ;
\end{tikzpicture}
.
\end{center}
All the above $4$-simplices have different $4$-volumes, tetrahedra volumes and triangle areas. To these one should also add $20$ more $4$-simplices, which are obtained by exchanging $a\leftrightarrow b$, i.e. by switching thick and thin lines. This gives a total of $40$ inequivalent isosceles $4$-simplices.

There are four different types of triangles that make up isosceles $4$-simplices. These are $(a,a,a)$, $(a,a,b)$, $(a,b,b)$ and $(b,b,b)$ triangles. Most of the $4$-simplices contain either three or sometimes all four types of triangles, making them unsuitable for satisfying the simplicity constraint, as explained in the main text. Nevertheless, there are five pairs of $4$-simplices which are made up only of two types of triangles. These are:
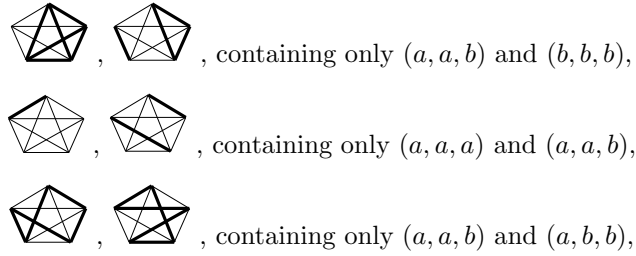
\begin{center}
\begin{tikzpicture}[yscale=1.5]
\draw[very thick] (-0.3,0) -- (0.3,0) ;
\draw[very thick] (-0.3,0) -- (0.5,0.3) ;
\draw[very thick] (-0.3,0) -- (0,0.5) ;
\draw[very thin] (-0.3,0) -- (-0.5,0.3) ;
\draw[very thick] (0.5,0.3) -- (0.3,0) ;
\draw[very thick] (0.3,0) -- (0,0.5) ;
\draw[very thin] (-0.5,0.3) -- (0.3,0) ;
\draw[very thick] (0.5,0.3) -- (0,0.5) ;
\draw[very thin] (-0.5,0.3) -- (0.5,0.3) ;
\draw[very thin] (-0.5,0.3) -- (0,0.5) ;
\end{tikzpicture}
,\ 
\begin{tikzpicture}[yscale=1.5]
\draw[very thin] (-0.3,0) -- (0.3,0) ;
\draw[very thin] (-0.3,0) -- (0.5,0.3) ;
\draw[very thin] (-0.3,0) -- (0,0.5) ;
\draw[very thick] (-0.3,0) -- (-0.5,0.3) ;
\draw[very thick] (0.5,0.3) -- (0.3,0) ;
\draw[very thick] (0.3,0) -- (0,0.5) ;
\draw[very thin] (-0.5,0.3) -- (0.3,0) ;
\draw[very thick] (0.5,0.3) -- (0,0.5) ;
\draw[very thin] (-0.5,0.3) -- (0.5,0.3) ;
\draw[very thin] (-0.5,0.3) -- (0,0.5) ;
\end{tikzpicture}
, containing only $(a,a,b)$ and $(b,b,b)$,
\end{center}
\begin{center}
\begin{tikzpicture}[yscale=1.5]
\draw[very thin] (-0.3,0) -- (0.3,0) ;
\draw[very thin] (-0.3,0) -- (0.5,0.3) ;
\draw[very thin] (-0.3,0) -- (0,0.5) ;
\draw[very thin] (-0.3,0) -- (-0.5,0.3) ;
\draw[very thin] (0.5,0.3) -- (0.3,0) ;
\draw[very thin] (0.3,0) -- (0,0.5) ;
\draw[very thin] (-0.5,0.3) -- (0.3,0) ;
\draw[very thin] (0.5,0.3) -- (0,0.5) ;
\draw[very thin] (-0.5,0.3) -- (0.5,0.3) ;
\draw[very thick] (-0.5,0.3) -- (0,0.5) ;
\end{tikzpicture}
,\ 
\begin{tikzpicture}[yscale=1.5]
\draw[very thin] (-0.3,0) -- (0.3,0) ;
\draw[very thin] (-0.3,0) -- (0.5,0.3) ;
\draw[very thin] (-0.3,0) -- (0,0.5) ;
\draw[very thin] (-0.3,0) -- (-0.5,0.3) ;
\draw[very thin] (0.5,0.3) -- (0.3,0) ;
\draw[very thin] (0.3,0) -- (0,0.5) ;
\draw[very thick] (-0.5,0.3) -- (0.3,0) ;
\draw[very thick] (0.5,0.3) -- (0,0.5) ;
\draw[very thin] (-0.5,0.3) -- (0.5,0.3) ;
\draw[very thin] (-0.5,0.3) -- (0,0.5) ;
\end{tikzpicture}
, containing only $(a,a,a)$ and $(a,a,b)$,
\end{center}
\begin{center}
\begin{tikzpicture}[yscale=1.5]
\draw[very thin] (-0.3,0) -- (0.3,0) ;
\draw[very thin] (-0.3,0) -- (0.5,0.3) ;
\draw[very thick] (-0.3,0) -- (0,0.5) ;
\draw[very thick] (-0.3,0) -- (-0.5,0.3) ;
\draw[very thick] (0.5,0.3) -- (0.3,0) ;
\draw[very thin] (0.3,0) -- (0,0.5) ;
\draw[very thick] (-0.5,0.3) -- (0.3,0) ;
\draw[very thick] (0.5,0.3) -- (0,0.5) ;
\draw[very thin] (-0.5,0.3) -- (0.5,0.3) ;
\draw[very thin] (-0.5,0.3) -- (0,0.5) ;
\end{tikzpicture}
,\ 
\begin{tikzpicture}[yscale=1.5]
\draw[very thick] (-0.3,0) -- (0.3,0) ;
\draw[very thick] (-0.3,0) -- (0.5,0.3) ;
\draw[very thin] (-0.3,0) -- (0,0.5) ;
\draw[very thin] (-0.3,0) -- (-0.5,0.3) ;
\draw[very thin] (0.5,0.3) -- (0.3,0) ;
\draw[very thick] (0.3,0) -- (0,0.5) ;
\draw[very thin] (-0.5,0.3) -- (0.3,0) ;
\draw[very thin] (0.5,0.3) -- (0,0.5) ;
\draw[very thick] (-0.5,0.3) -- (0.5,0.3) ;
\draw[very thick] (-0.5,0.3) -- (0,0.5) ;
\end{tikzpicture}
, containing only $(a,a,b)$ and $(a,b,b)$,
\end{center}
and their $a\leftrightarrow b$ duals (the third pair is self-dual). Of these pairs, one can use the first pair (or their dual) to construct a triangulation which features $3$-dimensional hypersurfaces made of equilateral tetrahedra, such that these hypersurfaces are separated by ``equal distances'' everywhere. Thus one can establish the triangulation-induced foliation of spacetime into space and time. This property is exploited by the CDT approach to quantum gravity. The remaining pairs above lack this property, because they mix edges of size $a$ and $b$ in a way that does not admit a nice $3+1$ foliation.

\section{\label{ApendiksC}Asymptotics of the weak-constrained Regge action}

Here we give the proof of the statement from the main text that the action (\ref{ReggeDejstvoSaWeakConstraints}) becomes equal to the proper Regge action in the asymptotic limit of large edge lengths. As noted below equation (\ref{ReggeDejstvoSaWeakConstraints}), the limit is defined by rescaling all edge lengths $l_{\epsilon}$ by a common factor $k$, and then expanding the resulting action into the asymptotic series when $k\to\infty$. Therefore, we start from (\ref{ReggeDejstvoSaWeakConstraints}) evaluated at $kl_{\epsilon}$,
$$
S_R(kl) \equiv \sum_{\Delta\in T} \left\lfloor \frac{1}{\gamma l_p^2} A_H(kl_{\epsilon\in\Delta}) \right\rfloor \delta_{\Delta}(kl)\,,
$$
and expand it into the asymptotic series as follows. First we use the fact that for any $x\in\realni$, the floor function can be written in the form
$$
\left\lfloor x \right\rfloor = x - R(x)\,,
$$
where the remainder $R(x)$ is always such that $ R(x) \in [0,1)$. Taking $x$ to be $A_H(kl)/\gamma l_p^2$ and applying the above formula to our action, the first term can be recognized as the proper Regge action (\ref{ProperReggeAction}), while the second term contains the reminder. We thus obtain
\begin{equation} \label{AsimptotskiRazvojWeakReggeDejstvaUsputnaJna}
S_R(kl) = S_{\rm Regge}(kl) - \sum_{\Delta\in T} \tilde{R}(kl) \delta_{\Delta}(kl)\,,
\end{equation}
where
$$
\tilde{R}(kl) = R\left( \frac{1}{\gamma l_p^2} A_H(kl) \right)\,,
$$
and therefore
\begin{equation} \label{KodomenRemaindera}
0 \leq \tilde{R}(kl) < 1
\end{equation}
for every $kl_{\epsilon} \in \realni$.

Next we need to recall from geometry that the deficit angles are homogeneous functions of order zero in the edge lengths, while Heron formula for triangle area is a homogeneous function of order two in the edge lengths,
$$
\delta_{\Delta}(kl) = \delta_{\Delta}(l)\,, \qquad A_H(kl) = k^2 A_H(l)\,.
$$
Therefore, the Regge action (\ref{ProperReggeAction}) scales as $S_{\rm Regge}(kl) = k^2 S_{\rm Regge}(l)$, so we can rewrite (\ref{AsimptotskiRazvojWeakReggeDejstvaUsputnaJna}) as
$$
S_R(kl) = S_{\rm Regge}(kl)\left[ 1 - \frac{1}{k^2 S_{\rm Regge}(l)} \sum_{\Delta\in T} \tilde{R}(kl) \delta_{\Delta}(l)\right] \,.
$$
Due to (\ref{KodomenRemaindera}), the second term on the right-hand side scales as $\cO(1/k^2)$ in the limit $k\to\infty$. Therefore, in this limit the action (\ref{ReggeDejstvoSaWeakConstraints}) does indeed become asymptotically equal to the Regge action (\ref{ProperReggeAction}),
$$
S_R(kl) = S_{\rm Regge}(kl)\left[ 1 - \cO\left(\frac{1}{k^2}\right) \right]\,, \qquad (k\to\infty) \,,
$$
as stated in the main text.

\section{\label{ApendiksD}Numerical analysis of simplicity constraint solutions}

In section \ref{SecSolutionToTheSimplicityConstraint}, we discussed the system of simplicity constraint equations (\ref{SistemJnaZaJedanSimplex}) for a single $4$-simplex, and claimed that the set of solutions is quite rich, based on the numerical Monte-Carlo analysis. Here we provide some details of that analysis.

We begin by noting that, for every triangle $\Delta$, its simplicity constraint equation (\ref{SimplicitySistemJna}) can be rewritten in the form
$$
|m_{\Delta}| = \frac{1}{\gamma l_p^2} A_H(l_1,l_2,l_3) = A_H\left( L_1, L_2, L_3 \right)\,,
$$
where $L_k = l_k / \sqrt{\gamma} l_p \in\realni$ are the dimensionless edge lengths of the triangle $\Delta$. This is due to the fact that the Heron formula for the triangle area is a homogeneous function of order two in the edge lengths. Therefore, if we label the vertices of a $4$-simplex as in the diagram
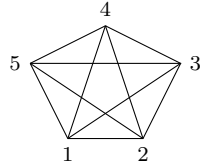
\begin{center}
\begin{tikzpicture}
\draw (-0.5,0) -- (0.5,0) ;
\draw (-0.5,0) -- (1,1) ;
\draw (-0.5,0) -- (-1,1) ;
\draw (-0.5,0) -- (0,1.5) ;
\draw (-1,1) -- (1,1) ;
\draw (-1,1) -- (0,1.5) ;
\draw (-1,1) -- (0.5,0) ;
\draw (1,1) -- (0,1.5) ;
\draw (1,1) -- (0.5,0) ;
\draw (0.5,0) -- (0,1.5) ;
\filldraw[black] (-0.5,0) circle (0pt) node[anchor=north] {\footnotesize 1};
\filldraw[black] (0.5,0) circle (0pt) node[anchor=north] {\footnotesize 2};
\filldraw[black] (1,1) circle (0pt) node[anchor=west] {\footnotesize 3};
\filldraw[black] (0,1.5) circle (0pt) node[anchor=south] {\footnotesize 4};
\filldraw[black] (-1,1) circle (0pt) node[anchor=east] {\footnotesize 5};
\end{tikzpicture}
\end{center}
we can write the simplicity constraint system of equations explicitly as follows:
\begin{equation} \label{NumerickiSistemJna}
\begin{array}{lcl}
|m_{123}| & = & A_H\left( L_{12}, L_{13}, L_{23} \right)\,, \\
|m_{124}| & = & A_H\left( L_{12}, L_{14}, L_{24} \right)\,, \\
|m_{125}| & = & A_H\left( L_{12}, L_{15}, L_{25} \right)\,, \\
|m_{134}| & = & A_H\left( L_{13}, L_{14}, L_{34} \right)\,, \\
|m_{135}| & = & A_H\left( L_{13}, L_{15}, L_{35} \right)\,, \\
|m_{145}| & = & A_H\left( L_{14}, L_{15}, L_{45} \right)\,, \\
|m_{234}| & = & A_H\left( L_{23}, L_{24}, L_{34} \right)\,, \\
|m_{235}| & = & A_H\left( L_{23}, L_{25}, L_{35} \right)\,, \\
|m_{245}| & = & A_H\left( L_{24}, L_{25}, L_{45} \right)\,, \\
|m_{345}| & = & A_H\left( L_{34}, L_{35}, L_{45} \right)\,. \\
\end{array}
\end{equation}
Here, each triangle and each edge is labeled with the vertices that belong to it.

The algorithm we used for the analysis of the solutions of the system (\ref{NumerickiSistemJna}) goes as follows. First, we randomly choose the ten integers $m_{\Delta}\in \{ 1,\dots, 50 \}$, where restricting to positive integers does not lead to loss of generality. Given this data, we proceed to look for all numerical solutions of (\ref{NumerickiSistemJna}) for the ten edge lengths $L_{\epsilon} \in [0,100]$. By randomly seeding $1000$ initial guesses for ten edges $L_{\epsilon}$ from the given domain, we employ the implementation of the Newton's method to find solutions of (\ref{NumerickiSistemJna}), up to machine precision. All found solutions are memorized and counted. For each solution, we calculate the Euclidean volume of the corresponding $4$-simplex, using the Cayley-Menger determinant formula
\begin{equation} \label{CayleyDet}
V_E^2 = - \frac{1}{9216} \left| 
\begin{array}{cccccc}
0 & 1 & 1 & 1 & 1 & 1 \\
1 & 0        & L_{12}^2 & L_{13}^2 & L_{14}^2 & L_{15}^2 \\
1 & L_{12}^2 & 0        & L_{23}^2 & L_{24}^2 & L_{25}^2 \\
1 & L_{13}^2 & L_{23}^2 & 0        & L_{34}^2 & L_{35}^2 \\
1 & L_{14}^2 & L_{24}^2 & L_{34}^2 & 0        & L_{45}^2 \\
1 & L_{15}^2 & L_{25}^2 & L_{35}^2 & L_{45}^2 & 0      \\
\end{array}
 \right|\,.
\end{equation}
Only solutions having different $4$-volumes are retained. After the seed of $1000$ initial edge lengths has been exhausted, the algorithm returns to the beginning, randomly choosing another set of integers $m_{\Delta}$, and repeats the calculation.

One should note that the Newton's method in general does not guarantee to find all possible solutions in a given range, since some of the solutions may only be reached by choosing a seed outside the given domain. Nevertheless, for our purposes it is not necessary to find all solutions, but merely some of them, since our goal is to demonstrate that the system (\ref{NumerickiSistemJna}) generically has many solutions.

The results of the analysis are as follows. We have counted the number $M$ of randomly chosen integers $m_{\Delta}$ that have $n$ different solutions for the ten edge lengths $L_{\epsilon}$, such that the $4$-volume of each solution is different from the others, and such that the $4$-simplex can be embedded into a $4$-dimensional space with Lorentz signature. The latter condition is achieved if the Cayley-Menger determinant (\ref{CayleyDet}) is negative. The function $M(n)$ thus gives us the distribution of choices of $m_{\Delta}$ over the number of distinct Lorentzian solutions for $L_{\epsilon}$ of the simplicity constraint system (\ref{NumerickiSistemJna}). Normalized over $1000$ choices of $m_{\Delta}$, the obtained distribution is:
$$
\begin{array}{|c||c|c|c|c|c|c|c|c|c|c|c|c|c|c|c|c|c|c|c|c|c|} \hline
n    & \;\;1\;\; & \;\;2\;\; & \;\;3\;\; & \;\;4\;\; & \;\;5\;\; & \;\;6\;\; & \;\;7\;\; & \;\;8\;\; & \;\;9\;\; & \;10\; & \;11\; & \;12\; & \;13\; & \;14\; & \;15\; & \;16\; & \;17\; & \;18\; & \;19\; & \;20\; & \;21\; \\ \hline
M(n) & 0 & 5 & 15 & 37 & 51 & 87 & 106 & 137 & 124 & 112 & 107 & 89 & 64 & 28 & 16 &  8 &  7 &  4 &  2 &  1 &  0 \\ \hline
\end{array}
$$
This can be represented as a histogram, in the following way:
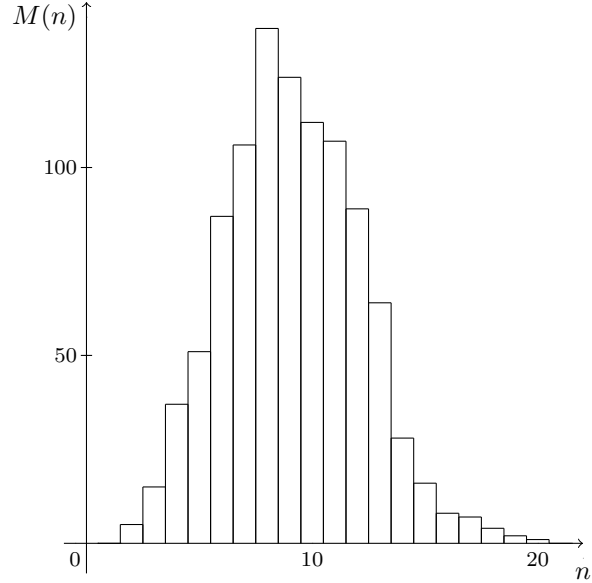
\begin{center}
\begin{tikzpicture}[xscale=0.3, yscale=0.2]
\draw[thin,->] (-1,0) -- (22,0);
\draw[thin,->] (0,-2) -- (0,36);
\draw[thin] (-0.25,12.5) -- (0.25,12.5);
\draw[thin] (-0.25,25) -- (0.25,25);
\filldraw[black] (10,0) circle (0pt) node[anchor=north] {\footnotesize 10};
\filldraw[black] (20,0) circle (0pt) node[anchor=north] {\footnotesize 20};
\filldraw[black] (0,12.5) circle (0pt) node[anchor=east] {\footnotesize 50};
\filldraw[black] (0,25) circle (0pt) node[anchor=east] {\footnotesize 100};
\filldraw[black] (-0.5,0) circle (0pt) node[anchor=north] {\footnotesize 0};
\filldraw[black] (22,-1) circle (0pt) node[anchor=north] {$n$};
\filldraw[black] (0,35) circle (0pt) node[anchor=east] {$M(n)$};
\draw  (1.5,0) --  (1.5,0) ;
\draw  (2.5,0) --  (2.5,1.25) ;
\draw  (3.5,0) --  (3.5,3.75) ;
\draw  (4.5,0) --  (4.5,9.25) ;
\draw  (5.5,0) --  (5.5,12.75) ;
\draw  (6.5,0) --  (6.5,21.75) ;
\draw  (7.5,0) --  (7.5,26.5) ;
\draw  (8.5,0) --  (8.5,34.25) ;
\draw  (9.5,0) --  (9.5,31) ;
\draw (10.5,0) -- (10.5,28) ;
\draw (11.5,0) -- (11.5,26.75) ;
\draw (12.5,0) -- (12.5,22.25) ;
\draw (13.5,0) -- (13.5,16) ;
\draw (14.5,0) -- (14.5,7) ;
\draw (15.5,0) -- (15.5,4) ;
\draw (16.5,0) -- (16.5,2) ;
\draw (17.5,0) -- (17.5,1.75) ;
\draw (18.5,0) -- (18.5,1) ;
\draw (19.5,0) -- (19.5,0.5) ;
\draw (20.5,0) -- (20.5,0.25) ;
\draw (21.5,0) -- (21.5,0) ;
\draw  (0.5,0) --  (0.5,0) ;
\draw  (1.5,0) --  (1.5,1.25) ;
\draw  (2.5,0) --  (2.5,3.75) ;
\draw  (3.5,0) --  (3.5,9.25) ;
\draw  (4.5,0) --  (4.5,12.75) ;
\draw  (5.5,0) --  (5.5,21.75) ;
\draw  (6.5,0) --  (6.5,26.5) ;
\draw  (7.5,0) --  (7.5,34.25) ;
\draw  (8.5,0) --  (8.5,31) ;
\draw  (9.5,0) --  (9.5,28) ;
\draw (10.5,0) -- (10.5,26.75) ;
\draw (11.5,0) -- (11.5,22.25) ;
\draw (12.5,0) -- (12.5,16) ;
\draw (13.5,0) -- (13.5,7) ;
\draw (14.5,0) -- (14.5,4) ;
\draw (15.5,0) -- (15.5,2) ;
\draw (16.5,0) -- (16.5,1.75) ;
\draw (17.5,0) -- (17.5,1) ;
\draw (18.5,0) -- (18.5,0.5) ;
\draw (19.5,0) -- (19.5,0.25) ;
\draw (20.5,0) -- (20.5,0) ;
\draw  (0.5,0) --  (1.5,0) ;
\draw  (1.5,1.25) --  (2.5,1.25) ;
\draw  (2.5,3.75) --  (3.5,3.75) ;
\draw  (3.5,9.25) --  (4.5,9.25) ;
\draw  (4.5,12.75) --  (5.5,12.75) ;
\draw  (5.5,21.75) --  (6.5,21.75) ;
\draw  (6.5,26.5) --  (7.5,26.5) ;
\draw  (7.5,34.25) --  (8.5,34.25) ;
\draw  (8.5,31) --  (9.5,31) ;
\draw  (9.5,28) -- (10.5,28) ;
\draw (10.5,26.75) -- (11.5,26.75) ;
\draw (11.5,22.25) -- (12.5,22.25) ;
\draw (12.5,16) -- (13.5,16) ;
\draw (13.5,7) -- (14.5,7) ;
\draw (14.5,4) -- (15.5,4) ;
\draw (15.5,2) -- (16.5,2) ;
\draw (16.5,1.75) -- (17.5,1.75) ;
\draw (17.5,1) -- (18.5,1) ;
\draw (18.5,0.5) -- (19.5,0.5) ;
\draw (19.5,0.25) -- (20.5,0.25) ;
\draw (20.5,0) -- (21.5,0) ;
\end{tikzpicture}
\end{center}

As we can see, for a random choice of ten $m_{\Delta}$, the simplicity constraint system (\ref{NumerickiSistemJna}) is most likely to have somewhere between $6$ and $12$ solutions with Lorentzian signature. No choices of $m_{\Delta}$ have been found to have only one, or more than $20$ solutions. The record number of $20$ solutions have been found for the following choice:
\begin{equation} \label{MaksimalniEmovi}
\begin{array}{lcl}
m_{123} = 28 \,, & & m_{145} = 27 \,, \\
m_{124} = 42 \,, & & m_{234} = 42 \,, \\
m_{125} = 43 \,, & & m_{235} = 38 \,, \\
m_{134} = 32 \,, & & m_{245} = 28 \,, \\
m_{135} = 26 \,, & & m_{345} = 24 \,. \\
\end{array}
\end{equation}
The corresponding solutions for $L_{\epsilon}$, as well as the values of the $V_E^2$, are as follows (note that these results have been obtained with $64$-bit machine precision, but are quoted here only up to three decimal places, for brevity):
\begin{widetext}
$$
\begin{array}{|c||r@{.}l|r@{.}l|r@{.}l|r@{.}l|r@{.}l|r@{.}l|r@{.}l|r@{.}l|r@{.}l|r@{.}l|r@{.}l|} \hline
\alpha
 & \multicolumn{2}{c|}{L_{12}}
 & \multicolumn{2}{c|}{L_{13}}
 & \multicolumn{2}{c|}{L_{14}}
 & \multicolumn{2}{c|}{L_{15}}
 & \multicolumn{2}{c|}{L_{23}}
 & \multicolumn{2}{c|}{L_{24}}
 & \multicolumn{2}{c|}{L_{25}}
 & \multicolumn{2}{c|}{L_{34}}
 & \multicolumn{2}{c|}{L_{35}}
 & \multicolumn{2}{c|}{L_{45}}
 & \multicolumn{2}{c|}{V_E^2} \\ \hline\hline
  1 &  9 & 188 &  6 & 866 &  9 & 647 & 15 & 263 &  8 & 572 & 15 & 298 &  9 & 790 & 10 & 301 &  9 & 907 &  7 & 216 & -96358 & 047 \\ \hline 
  2 &  9 & 751 &  6 & 180 & 15 & 423 &  9 & 195 &  9 & 420 &  9 & 135 & 15 & 179 & 11 & 610 &  8 & 769 &  7 & 755 & -83095 & 409 \\ \hline 
  3 &  9 & 841 &  6 & 254 & 15 & 930 & 10 & 998 &  9 & 213 &  9 & 267 &  9 & 294 & 11 & 839 & 15 & 816 &  6 & 433 & -138374 & 305 \\ \hline 
  4 &  9 & 936 &  6 & 799 &  9 & 492 & 10 & 013 &  8 & 329 & 10 & 153 &  9 & 948 & 12 & 365 & 15 & 309 &  5 & 825 & -14626 & 743 \\ \hline 
  5 &  9 & 995 &  6 & 191 & 10 & 364 &  9 & 425 &  9 & 251 &  9 & 278 & 10 & 576 & 11 & 682 &  8 & 615 & 18 & 949 & -29435 & 832 \\ \hline 
  6 & 10 & 010 &  6 & 334 & 10 & 261 & 15 & 153 &  8 & 992 &  9 & 341 &  8 & 940 & 12 & 964 & 10 & 411 &  6 & 579 & -65535 & 329 \\ \hline 
  7 & 10 & 012 &  6 & 292 & 10 & 173 & 15 & 024 &  9 & 063 &  9 & 409 &  8 & 896 & 11 & 887 & 10 & 365 &  6 & 576 & -43342 & 444 \\ \hline 
  8 & 10 & 152 &  7 & 352 &  8 & 705 &  9 & 266 &  7 & 644 & 15 & 291 & 10 & 615 & 11 & 388 & 15 & 097 &  6 & 465 & -70393 & 956 \\ \hline 
  9 & 10 & 362 & 18 & 623 &  9 & 328 & 10 & 066 &  9 & 212 &  9 & 937 &  9 & 522 & 10 & 528 &  9 & 378 & 18 & 492 & -428991 & 593 \\ \hline 
 10 & 10 & 669 & 18 & 040 &  8 & 539 &  8 & 688 &  8 & 425 & 10 & 780 & 16 & 077 & 10 & 868 & 10 & 257 &  6 & 831 & -184052 & 701 \\ \hline 
 11 & 10 & 809 &  6 & 742 &  9 & 996 & 15 & 572 & 15 & 987 &  8 & 991 &  8 & 363 & 10 & 156 & 10 & 273 &  7 & 101 & -121442 & 024 \\ \hline 
 12 & 10 & 888 & 17 & 907 &  8 & 255 &  8 & 851 &  8 & 106 & 15 & 831 & 10 & 485 & 11 & 059 &  9 & 978 &  6 & 932 & -139620 & 742 \\ \hline 
 13 & 11 & 257 &  5 & 519 & 16 & 580 & 10 & 493 & 10 & 168 &  8 & 263 &  8 & 653 & 13 & 213 &  9 & 466 &  7 & 365 & -7145 & 581 \\ \hline 
 14 & 11 & 451 &  6 & 478 & 10 & 104 &  8 & 715 & 16 & 443 &  8 & 608 & 10 & 288 & 10 & 800 &  8 & 600 & 17 & 808 & -162923 & 501 \\ \hline 
 15 & 11 & 518 &  6 & 032 & 15 & 321 &  9 & 311 & 15 & 853 &  7 & 551 &  9 & 551 & 11 & 737 &  8 & 979 &  7 & 573 & -67437 & 306 \\ \hline 
 16 & 11 & 710 &  7 & 059 &  9 & 385 &  7 & 464 &  8 & 084 & 19 & 156 & 12 & 714 & 13 & 118 &  9 & 413 &  7 & 386 & -2106 & 96 \\ \hline 
 17 & 11 & 875 & 19 & 723 &  8 & 718 &  9 & 650 &  8 & 671 &  9 & 798 &  9 & 092 & 12 & 076 & 10 & 768 &  6 & 333 & -26923 & 715 \\ \hline 
 18 & 13 & 967 &  6 & 434 & 10 & 134 &  8 & 083 & 19 & 418 &  8 & 362 & 10 & 683 & 13 & 002 & 10 & 254 &  6 & 697 & -4065 & 232 \\ \hline 
 19 & 14 & 069 &  7 & 238 &  8 & 847 &  9 & 425 &  8 & 957 &  9 & 618 &  9 & 214 & 11 & 595 & 15 & 150 &  6 & 327 & -68517 & 683 \\ \hline 
 20 & 14 & 739 &  6 & 577 &  9 & 842 &  8 & 854 & 10 & 111 &  8 & 807 &  9 & 966 & 10 & 986 &  8 & 320 & 17 & 665 & -119422 & 745 \\ \hline
\end{array}
$$
\end{widetext}
One can see that all solutions describe $4$-simplices with different $4$-volumes. Moreover, it is straightforward to verify, using Heron's formula, that for each choice of edge lengths, all ten triangle areas are integers (\ref{MaksimalniEmovi}).

Finally, we should note that the only purpose of the results obtained in this analysis is to illustrate that the set of solutions of the system (\ref{SistemJnaZaJedanSimplex}) is indeed quite rich, as claimed in the main text.

\end{document}